\documentclass[a4paper,10pt]{article}

 \usepackage[cp1251]{inputenc}
 \usepackage{cite}

 \usepackage{graphicx}

\usepackage{cite,amsmath,amsfonts,amsthm,fullpage}
\usepackage{youngtab}

\newcommand{\bpow}{\mathbf{p}}

\newcommand{\bt}{\mathbf{t}}

\newcommand{\gF}{\mathfrak{F}}
\newcommand{\gC}{\mathfrak{C}}
\newcommand{\gJ}{\mathfrak{J}}

\newcommand{\f}{\textsc{f}}
\newcommand{\e}{\textsc{e}}
\newcommand{\V}{\textsc{v}}

\def\q{\texttt{q}}

\theoremstyle{plain}

\newtheorem{Theorem}{Theorem}
\newtheorem{Lemma}{Lemma}
\newtheorem{Proposition}{Proposition}
\newtheorem{Corollary}{Corollary}
\newtheorem{Remark}{Remark}

\theoremstyle{remark}

\def\g{\Gamma}

\def\bp{\begin{Proposition}}
\def\ep{\end{Proposition}}
\def\bc{\begin{Corollary}}
\def\ec{\end{Corollary}}
\def\bl{\begin{Lemma}}
\def\el{\end{Lemma}}
\def\be{\begin{equation}}
\def\ee{\end{equation}}
\def\br{\begin{Remark}\rm\small}
\def\er{\end{Remark}}
\def\brs{\begin{remarks}.\\ \rm\
\begin{enumerate}}
\def\ers{\end{enumerate}\end{remarks}}
\def\bea{\begin{eqnarray}}
\def\eea{\end{eqnarray}}

\def\det{\mathrm {det}}

\def\ln{\mathrm {ln}}

\def\res{\mathop{\mathrm {res}}\limits_}

\def\ln{\mathrm {ln}}

\def\res{\mathop{\mathrm {res}}\limits}

\def\&{&{\hskip -20pt}}

\usepackage[usenames,dvipsnames]{color}
\usepackage{ulem}

\begin{document}

\author{ Sergey M. Natanzon\thanks{National Research University Higher School of Economics, Moscow, Russia; 
Institute for Theoretical and Experimental Physics, Moscow, Russia;
email: natanzons@mail.ru} \and Aleksandr Yu.
Orlov\thanks{Institute of Oceanology, Nahimovskii Prospekt 36,
Moscow 117997, Russia, and National Research University Higher School of Economics,
International Laboratory of Representation
Theory and Mathematical Physics,
20 Myasnitskaya Ulitsa, Moscow 101000, Russia, email: orlovs@ocean.ru
}}
\title{Hurwitz numbers from matrix integrals over Gaussian measure}

\maketitle

\begin{abstract}

\noindent 

We explain how Gaussian integrals over ensemble of complex matrices with source matrices generate Hurwitz numbers 
of the most general type, namely, Hurwitz numbers with arbitrary orientable or non-orientable base surface and
arbitrary profiles at branch points. We use the Feynman diagram approach.
The connections with topological theories and also with certain classical and quantum
integrable theories in particular with Witten's description of two-dimensional gauge theory are shown.

\end{abstract}

\bigskip


\textbf{2010 Mathematic Subject Classification:} 05A15, 14N10, 17B80, 35Q51, 35Q53, 35Q55, 37K20, 37K30,

\bigskip

\qquad\qquad\qquad\qquad\qquad  In memory of Boris Dubrovin, a great mathematician and friend.

\bigskip

\section{Introduction}

There is a lot of literature on Hurwitz numbers, on the connections of Hurwitz numbers with topological 
field theories
\cite{Dijkgraaf},\cite{D2},\cite{N}, and with integrable systems 
\cite{Okounkov-2000},\cite{Okounkov-Pand-2006},\cite{Goulden-Jackson-2008},\cite{MM4},  
\cite{MM5},\cite{Takasaki-Hurwitz},\cite{AMMN-2014},\cite{Goulden-Paquet-Novak},
\cite{HO-2014},\cite{NO-LMP},\cite{GayPakettHarnad2},\cite{{NatanzonZabrodin}}
(see reviews in \cite{KazarianLando},\cite{HarnadOverview}),
and also on relations of Hurwitz numbers to matrix models \cite{MelloKochRamgoolam},\cite{Alexandrov},
\cite{Zog},\cite{GGPN},\cite{GayPakettHarnad1},\cite{NO-2014},\cite{Chekhov-2014},\cite{O-TMP-2017},\cite{O-2017}.
The papers most closely related to the present work are \cite{ChekhovAmbjorn} and \cite{Orlov-chord}.

Here, we consider three tasks. The first is to present models that allow the generation of Hurwitz numbers
with arbitrary given sets of profiles at isolated points, that describe the enumeration of non-isomorphic
covering maps of any closed surface. 
The second is to connect such models with topological field theories.
Finally, we look at relations between Hurwitz numbers and quantum integrable systems.
Such a relation was noticed by Dubrovin in connection with the quantum dispersionless KdV equation. The
second connection with integrable systems is that the correlation functions of two-dimensional quantum
gauge theory (2D Yang-Mills theory), defined on a closed orientable or non-orientable surface found by
E.Witten [99], can also be considered as generating functions for Hurwitz numbers.

In Section \ref{definitions} we give necessary definitions and present known facts about
Hurwitz numbers, topological field theories and relations with representations of symmetric group.
We collect the necessary consequences of axioms of the related topological field theory in 
the form of Proposition \ref{Hurwitz-down-Lemma}. The notion of independent Ginibre
ensembles is explained.

Section \ref{Geom comb Hurwitz} is central. 
There, in the case of orientable surfaces,
we present another way of determining the Hurwitz numbers, namely, as an enumeration of the number of 
possible ways to glue surfaces from sets of polygons.
Polygons are constructed from \textit{basic polygons} using Young diagrams attached to them.
Gluing the basic polygons leads to the formation of a base surface.
The enumeration of possible ways of gluing is determined using Wick's rule, applicable to Gaussian integrals.
Note that the Wick rule is widely used by physicists to interpret various matrix integrals.
It was first used by t'Hooft \cite{t'Hooft} for QCD matrix models, and then it was successfully used
for various problems in physics and mathematical physics, see \cite{Itzykson-Zuber}, \cite{BrezinKazakov}, 
\cite{Kazakov-ZinnJ}, \cite{Chekhov-2014} and many others.
For our purpose, we need a model of independent Ginibre ensembles with two sets of matrices: a set that we 
call the set of random
matrices and a set, which we call the set of source matrices. The last one is needed to get
specific Hurwitz numbers, see Theorem \ref{Theorem}. Let us note, that
Ginibre ensembles are popular models to study problems in quantum chaos \cite{FyodorovSommers} and
information transfer problems \cite{AkStrahov}, \cite{Alfano}.

To obtain Hurwitz numbers in case of non-orientable surface, we modify the measure of integration, 
see Theorem \ref{integral=handle-cuts} in Subsection \ref{non-orientable}, and then we
the axioms of Hurwitz topological field theory. To modify the measure we use tau functions.
 The tau function is the central object in the theory of classical integrable models.

In Section \ref{Hurwitz-quantum-models}, the Hurwitz numbers are related to specific quantum solvable models
of two different types, considered respectively in subsections \ref{Dubr-quant} and \ref{Witten-quant}.
In Subsection \ref{Dubr-quant}, we recall Dubrovin's observation that completed cycles (which were 
introduced into the theory of Hurwitz numbers by Okounkov in \cite{Okounkov-Pand-2006}))
 turn out to be the eigenvalues of quantum Hamiltonian operators, which are the  Hamiltonians
of "higher" quantum dispersionless KdV equations.
 We study this relation using Jucys-Murphy elements and a Toda lattice analogue of the dispersionless KdV equations.
We consider the action of such a Toda lattice Hamiltonian on our matrix models. This action generates the 
additional dependence of Hurwitz numbers on  the completed cycles introduced by Okounkov 
\cite{OkounkovVershik}. Next, we select all such 
integrals that can be equated to tau functions, see Propositions \ref{Prop-Jucys-Murphy-tau-TL} and 
\ref{Prop-Jucys-Murphy-tau-BKP} (such tau functions describe the lattice of solitons in 
classical integrable systems).

Subsection \ref{Witten-quant} shows that the partition function of the two-dimensional quantum gauge theory
with fixed holonomy around marked points, presented by Witten in \cite{Witten} can be considered as
integral over complex matrices with sources, and this partition function also generates Hurwitz numbers. In this case, 
we consider the integrands that are tau-functions of a special family found in \cite{KMMM},\cite{OS-TMP}.

\section{Definitions and a review of known results \label{definitions}}

\subsection {Definition of Hurwitz Numbers.}

\vspace {2ex}

Hurwitz number is a weighted number of branched coverings of a surface with a prescribed topological
type of critical values.  
Hurwitz numbers of oriented surfaces without boundaries were introduced by Hurwitz
at the end of the 19th century.
Later it turned out that they are closely related to the study of the module spaces of Riemann
surfaces \cite{ELSV}, to the integrable systems \cite{Okounkov-2000}, to modern models of mathematical physics
[matrix models], and to closed topological field theories \cite{Dijkgraaf}. In this paper we will consider only the
Hurwitz numbers over compact surfaces
without boundary. The definition and important
properties of Hurwitz numbers over arbitrary compact (possibly with boundary) surfaces were suggested in \cite{AN}.

Clarify the definition. Consider a branched covering $\varphi:P\rightarrow\Omega$ of degree $d$ over a compact surface
without boundary. In the neighborhood of each point $z\in P$, the map $\varphi$ is topologically equivalent to the
complex map $u\mapsto u^p$ in the neighborhood of $u=0\in\mathbb{C}$. The number $p=p(z)$ is called degree of the
covering $\varphi$ at the point $z$. The point $z\in P$ is called \textit{branch point} or  \textit{critical point}
if $p(z)\neq 1$. There is only a finite number of critical points. The image $\varphi(z)$ of any critical point is 
called \textit{critical value}.

Let us associate with a point $s\in\Omega$ all points  $z_1,\dots,z_\ell\in P$ such that $\varphi(z_i)=s$. Let
$p_1, \dots,p_\ell$ be the degrees of the map $\varphi$ at these points. Their sum $d=p_1 +\dots+p_\ell$ is equal
to the degree $d$ of $\varphi$. Thus, to each point $s\in S $ there corresponds a partition $d=p_1 +\dots+p_\ell$ of
the number $d$. 
Having ordered the degrees $ p_1 \geq \dots \geq p_\ell> 0 $ at each point $ s \in \Omega $, 
we can introduce the Young diagram $ \Delta^s = [p_1, \dots, p_ \ell ] $ of weight $ d $ with
$ \ell = \ell (\Delta^s) $ lines of length $ p_1 \dots, p_\ell $. 
This Young diagram $ \Delta^ s $ is called
\textit{topological type} of the value $ s $.
(We consider $ s $ to be a critical value if not all $ p_i $ are equal to $ 1 $.)

Let us note that the Euler characteristics $\e(P)$ and $\e(\Omega)$ of the surfaces $P$ and $\Omega$ are related
via the Riemann-Hurwitz relation:
\[
\e(P)=\e(\Omega)d +\sum\limits_{z\in P} \left(p(z)-1\right).
\]
or, which is the same,
\be\label{RHur}
\e(P)=\e(\Omega)d +\sum\limits_{i=1}^\f \left( \ell(\Delta^{s_i})-d\right).
\ee
where $s_1,\dots,s_{\f}$ are critical values.

We say that coverings $ \varphi_1: P_1 \rightarrow \Omega $ and $ \varphi_2: P_2 \rightarrow \Omega $ 
are \textit{equivalent}
if there exists a homeomorphism $ F: P_1 \rightarrow P_2 $ such that $ \varphi_1 = \varphi_2F $;
in case $\varphi_1 =\varphi_2$ the homeomorphism $F$ is called
\textit{automorphism of the covering}. Automorphisms of a covering $\varphi$ form the group $\texttt{Aut}(\varphi)$
of a finite order $|\texttt{Aut}(\varphi)|$. Equivalent coverings have isomorphic groups of automorphisms.

Let us choose points $ s_1, \dots, s_\f \in \Omega $ and Young diagrams $ \Delta^1, \dots, \Delta^\f $ of weight $ d $. 
Consider the set $ \Phi $ of all equivalence classes of the coverings for which $ s_1, \dots, s_\f $ is the set 
of all critical values, and $ \Delta^1, \dots, \Delta^\f $ are the topological types of these critical values.
Hereinafter, unless otherwise indicated we consider that the surface $ \Omega $ is connected.
\textit{Hurwitz number} is the number
\be\label{disconH} 
H_{\e(\Omega)}^d(\Delta^1,\dots,\Delta^\f)=\sum_{\varphi\in\Phi} 
\frac {1} {|\texttt{Aut} (\varphi)|}.
\ee
It is easy to prove that the Hurwitz number is independent of the positions of the points $s_1, \dots, s_\f$ 
on $\Omega$.
It depends only on the Young diagrams of $\Delta^1,\dots,\Delta^\f$ and the Euler characteristic $\e=\e(\Omega)$.

\vspace {2ex}

\subsection {Hurwitz Numbers and topological field theory.}

 The notion of closed topological field theory was proposed by
M. Atiayh in \cite{At}. 
This theory deals with closed oriented surfaces without boundary.
 We also fix a
finite-dimensional vector space $A$ with a basis $\alpha_1,\dots,\alpha_N$.
Consider now an arbitrary set of points $p_1,p_2,\dots,p_\f\in\Omega $. Let us assign a vector $a_i\in A$
to each point $p_i$. 

\medskip

\includegraphics[scale=0.5]{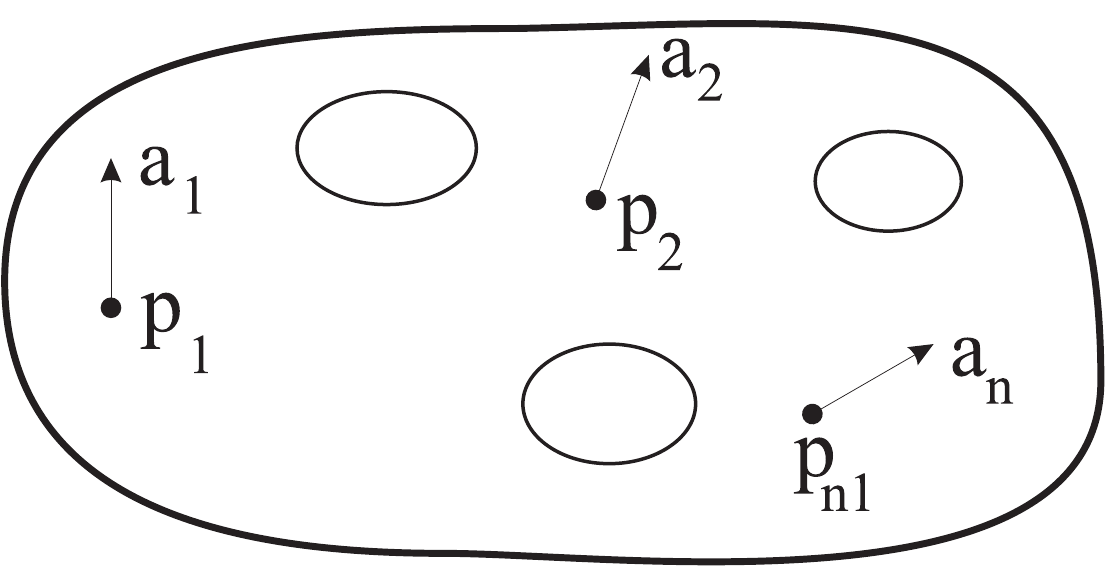}

\begin{figure} [tbph]
\caption {} \label{f1}
\end {figure}

\bigskip

A closed topological field theory attributes a number 
$ <a_1, a_2, \dots, a_\f>_{\Omega} $ called \textit{correlator}, to the dataset described above.
 We assume that the numbers $<a_1,a_2,\dots,a_\f>_{\Omega}$ are invariant
with respect to any autohomomorphism of the surface. Moreover, we assume that correlators
$ <a_1,a_2,\dots,a_\f>_{\Omega}$ generate a system of polylinear forms on $A$, which satisfies 
the axiom of nondegeneracy 
and the cutting axioms:

\textit{Axiom of Nondegeneracy} means that the bilinear form $ <a_1,a_2>_{\Omega}$ is nondegenerate. Let us denote 
the inverse matrix to $\left(<\alpha_i, \alpha_j>_{S^2} \right)_{1\le i, j\le N}$
by $ F^{\alpha_i,\alpha_j}_A$.

\textit{Axioms of cutting} describes the evolution of correlators
$ <a_1, a_2, \dots, a_\f>_{\Omega} $ as the result of the cutting the surface along a contour 
$ \gamma \subset \Omega $ with the subsequent contraction of each boundary contour
to a point.
Two possible topological types of contours give two axioms of cutting.
Suppose $\gamma$ splits the surface $\Omega$ into 2 surfaces $\Omega'$ and $\Omega''$ (Figure \ref{f2}).

\medskip

\includegraphics[scale=0.5]{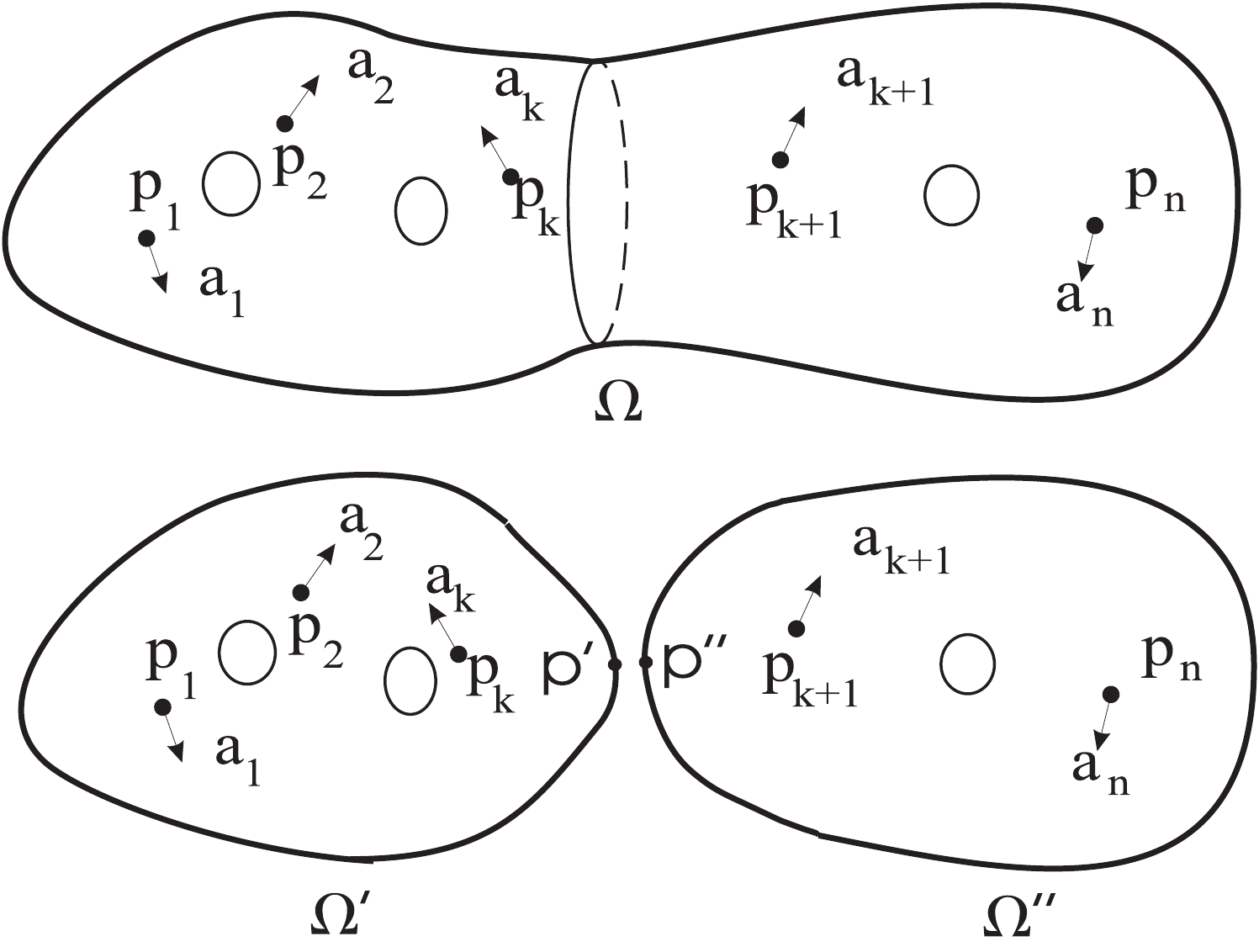}

\begin{figure} [tbph]
\caption{} \label{f2}
\end{figure}

Then
\be\label{cut}
<a_1,a_2,\dots,a_\f>_{\Omega}\, = \sum_{i, j}
<a_1,a_2,\dots,a_k,\alpha_i>_{\Omega'} F^{\alpha_i, \alpha_j}_A
<\alpha_j a_{k + 1},a_{k + 2}, ..., a_\f>_{\Omega ''}.
\ee

If $\gamma$ does not split $\Omega$ (figure\ref {f3}).
\medskip

\includegraphics[scale=0.5]{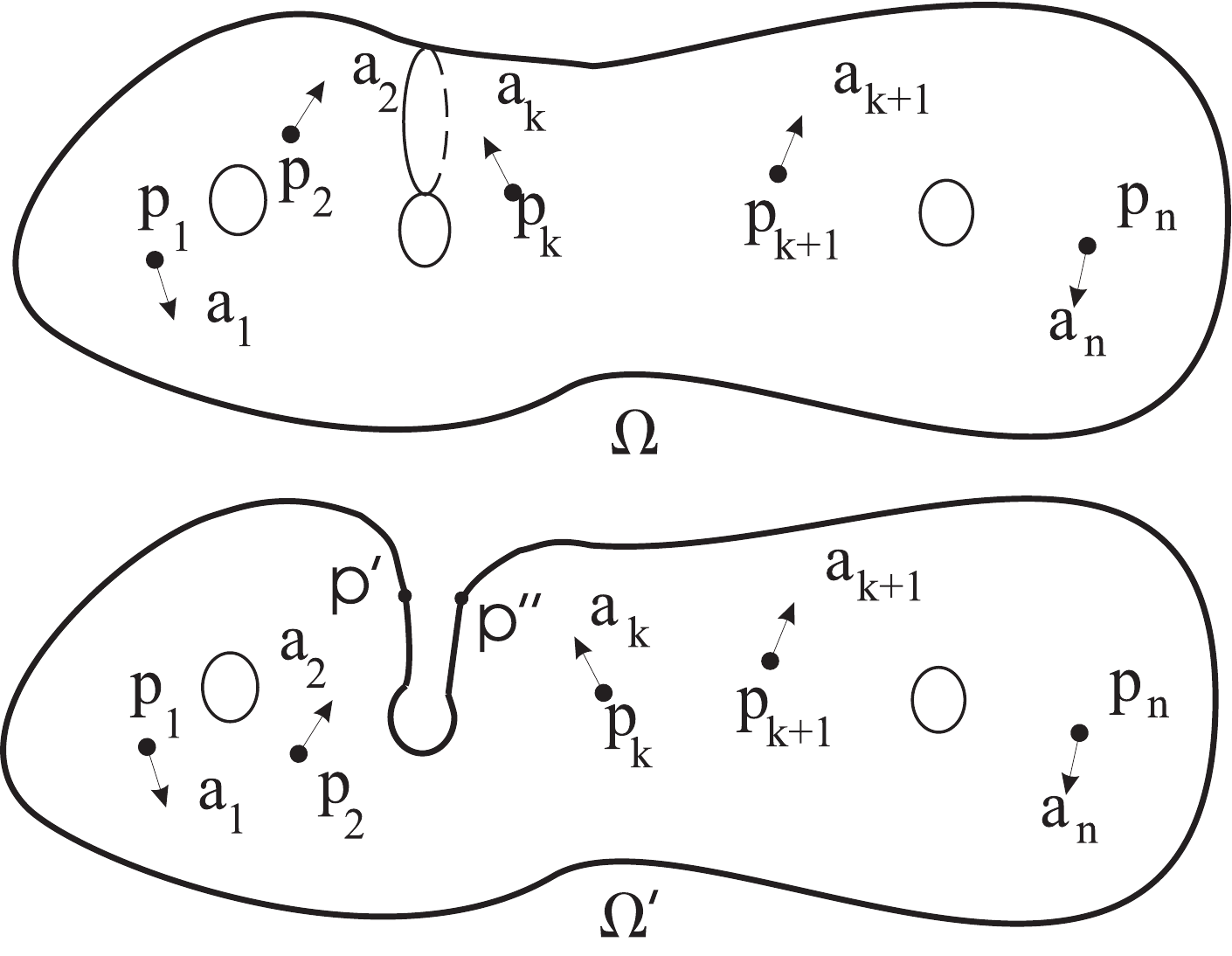}

\begin {figure} [tbph]
\caption{}\label {f3}
\end{figure}

then
\be\label{handle}
<a_1,a_2,\dots,a_\f>_{\Omega} \, =
\sum_{i,j} <a_1 a_2,\dots,a_\f,\alpha_i,\alpha_j>_{\Omega'} F^{\alpha_i,\alpha_j}_A.
\ee

The first corollary of the axioms of a topological field theory is the existance of an
\textit{algebra structure} on $A$. Namely, the multiplication is defined as
$ <a_1 a_2, \, a_3>_{S^2}\, = <a_1, \, a_2, \, a_3>_{S^2} $.
Thus, the structure constants for this algebra in the basis $\{\alpha_i\}$ are equal to
$c_{ij}^{k} = \sum_{s}<\alpha_i, \, \alpha_j, \, \alpha_s>_{S^2} F^{\alpha_s, \alpha_k}_A $.

The axiom of the cutting gives (Figure \ref{f4} )

\medskip

\includegraphics[scale=0.5]{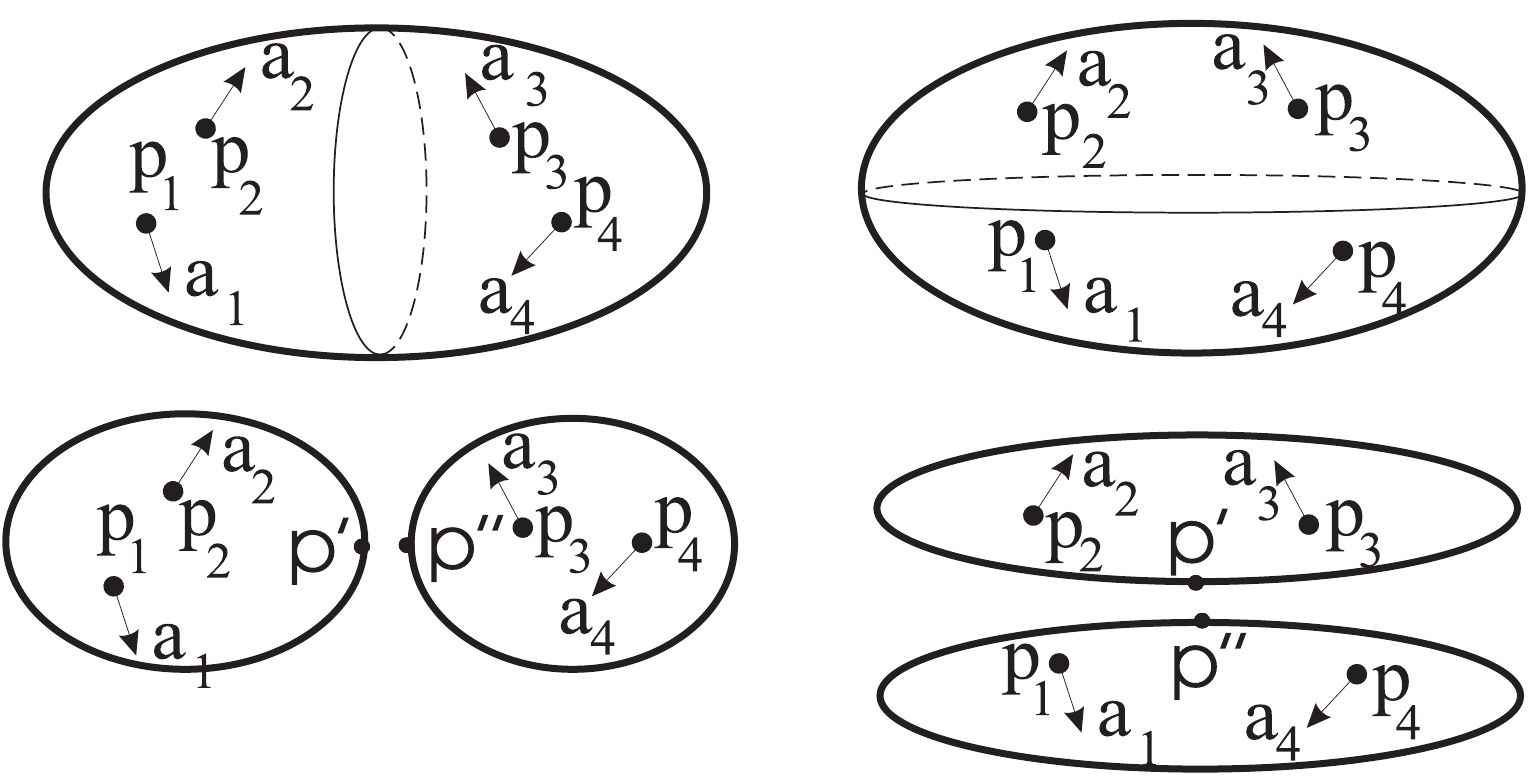}

\begin{figure} [tbph]
\caption{}\label {f4}
\end{figure}

$$
\sum_{i, j}
<a_1, a_2, \alpha_i>_{S^2} F_A^{\alpha_i, \alpha_j} <\alpha_j, a_3, a_4>_{S^2}\, =
$$
$$
= <a_1, a_2, a_3, a_4>_{S^2}\, =
$$
$$
\sum_{i, j}
<a_2, a_3, \alpha_i>_{S^2} F_A^{\alpha_i, \alpha_j} <\alpha_j, a_4, a_1>_{S^2}.
$$
Therefore, $\sum_{s,t}c_{ij}^{s}c_{sk}^{t}=\sum_{s,t}c_{jk}^{s}c_{si}^{t} $, that is, $A$ is an associative algebra.
Vector $\sum_{i}<\alpha_i>_{S^2}F^{\alpha_i, \alpha_j}_A \alpha_j$ is the unit of algebra $A$. The linear form
$l_A (a)= <a>_{S^2}$ generates non-degenerate invariant bilinear form $(a_1,a_2)_{A}=l(a_1a_2)= <a_1,a_2>_{S^2}$.
The topological invariance makes all marked points $p_i$ equivalent and, therefore, $A$ is a commutative algebra.

Thus, $A$ is a commutative Frobenius algebra \cite{F}, that is, an algebra with the unit and the invariant
nondegenerate scalar product generated by the linear functional $l_A$. We will call such pairs $(A,l_A)$
\textit{commutative Frobenius pairs}. Moreover, the described construction generates functor
$\mathcal{F}$ from the category of closed topological field theories to the category of commutative Frobenius pairs.

\begin{Theorem}\cite{D2} The functor $\mathcal{F}$ is the equivalence between the category of closed topological field
theories and the category of commutative Frobenius pairs.
\end{Theorem}

The structure of the Frobenius pair and the cutting axiom give the following explicit formula for correlators:
$$ <a_1,a_2,\dots,a_\f>_{\Omega} \,= l_A (a_1a_2\dots a_\f (K_A)^g)= <a_1,a_2,\dots,a_\f,(K_A)^g>_{S^2},$$ where
$K_A=\sum_{ij} F^{\alpha_i,\alpha_j}_A \alpha_i\alpha_j$  and $g$ is the genus of the surface $\Omega$.

\vspace {2ex}

\vspace{2ex}

An extension of topological field theories to non-orientable surfaces was proposed in \cite{AN}.
At the present paper, we consider surfaces without boundaries, and
we will call such theories \textit{closed Klein topological field theory}. A \textit{closed Klein  topological 
theory} is determined by the same scheme as
the closed one but here the correlator $<a_1,\dots,a_\f>_{\Omega}$ is defined for both orientable and 
non-orientable surfaces $\Omega$. In addition, each point of $p_i$ is supplied with the orientation of its 
small neighborhood, and the algebra $A$ is supplied with an involution $\star: A\rightarrow A$.
Besides, we assume that the change of the orientation of the neighborhood of a point $p_i$ changes the element
$a \in A$ placed at $p_i$ to $a^\star =\star(a)\in A$.

For non-orientable surfaces, in addition to the above-described cuts, there are 2 more types of cuts. 
The type depends on whether the contour is divided into one or two contours after the cut. In the first case, 
the cut is called the \textit{M\"{o}bius cut}, and in the second \textit{the Klein cut} (\cite{AN}).

Gluing along the Klein cut transforms correlators by the same rule as for non-dividing contours of closed topological
field theory:
\be\label{Klein-cut}
<a_1,a_2,\dots,a_\f>_{\Omega} = \sum_{i, j}<a_1, a_2,\dots,a_\f,\alpha_i,\alpha_j>_{\Omega'}F^{\alpha_i, \alpha_j}_A.
\ee

Gluing along the M\"{o}bius cut transforms correlators according to the rule
\be\label{Moebius-cut}
<a_1,a_2,\dots,a_\f>_{\Omega} =\sum \limits_{a_i}<a_{i_1}, \dots, a_{i_k}, a_i>_{\Omega'} D (a_i),
\ee
where the linear functional $D(a)= <a>_{\mathbb {RP}^2}: A \rightarrow \mathbb {K} $ is defined by the Klein
topological field theory for the real projective plane.

Let us denote by $U$ the element of algebra conjugate with respect to the (defined above) metric on $A$ to the linear
functional $D(a)$. Then it turns out that the quad $(A,l_A,U,\star)$ satisfies the following properties:

1) $ (A, l_A) $ is the commutative Frobenius pair;

2) the involution $ \star: A \rightarrow A $ generates an
automorphism of algebra;

3) $l_A(a^\star) = l_A(a)$;

4) $U^2 = F^{\alpha_i,\alpha_j}_A \alpha_i\alpha_j^\star$.

\begin{Theorem}\cite{AN} The functor $\mathcal{F}$ is extended to the category of closed Klein topological field
theories and defines an equivalence between the category of closed Klein topological field theories and the quadruplet
$(A,l_A,U,\star) $, with properties 1) -4). Moreover $<a_1, a_2, ..., a_\f>_{\Omega}\, =
l_A (a_1a_2 ... a_\f (U)^{2-e(\Omega)})$, where $e(\Omega)$ is the Euler characteristic of the surface $\Omega$.
\end{Theorem}

\vspace {2ex}

The closed Klein topological field theory can be extended to the Klein topological field theory, which includes
surfaces with a boundary \cite{AN}. Its algebraic description is given by a more complex algebraic structure,
called in \cite{AN1,AN2008} \textit{equipped Cardy-Frobenius algebra}. Moreover,  it was proved  in \cite{AN} that
the category of the Klein topological field theories is equivalent to the category of the equipped Cardy-Frobenius algebras.
This algebraic structure naturally arises also in the representation theory of finite groups \cite{LN}. Moreover,
Klein topological field theory can be continued on foams arising in string theory and algebraic geometry
\cite{N1}, \cite{MM3} \cite{CGN}, \cite{GN}.

Cohomological field theory is equivalent to a flat family of closed
topological field theories forming the Frobenius-Dubrovin manifold \cite{Dub, Man}.

Following the pattern of Klein topological field theory, we can construct  more general topological field theories
with values in  functors \cite{Nat}. For some special functor this theory gives the Kontsevich-Manin 
cohomological field theory and Gromov-Witten invariants \cite{KM94}.
Cohomological field theory is equivalent a flat family of closed 
topological field theories forming a Frobenius-Dubrovin manifold \cite{Dub,Man}. Dubrovin discovered that these 
manifolds play an important role in various branches of mathematics. Moreover, Frobenius-Dubrovin manifolds are
one-to-one 
correspond to quasihomogeneous solutions of the WDVV hierarchy of differential equations, which appear in theory 
of quantum gravity.

\vspace {3ex}

We now return to the Hurwitz numbers. Dijkgraaf \cite {Dijkgraaf} noticed that the Hurwitz numbers for coverings 
of degree $d$ generate certain
closed topological field theory with the vector space $Y_d$ generated by Young diagrams $\{\Delta\}$ of weight $d$.
Dijkgraaf considered correlators
\begin{equation}\label{corr}
< \Delta^1, \dots, \Delta^\f>_{\Omega} \,= H_{\e(\Omega)}^d (\Delta^1, \dots, \Delta^\f).
\end{equation}
and proved, that they satisfy all the axioms of closed topological field theory. If we continue this definition of
correlators for the case of non-orientable surfaces, we get closed Klein topological field theory (\cite{AN}). 
In this case, the operator $*$ is trivial.

Further, speaking of correlators , we will always mean the correlators $<*,\dots,*>_{\Omega}$  of the
\textit{Hurwitz topological field theory}. Put $<*,\dots,*>=<*,\dots,*>_{S^2}$. The linear functional $<\Delta>$ 
on $Y_d$ is equal to $\frac{1}{d!}$ for $\Delta=[1,\dots,1]$ and to
$0$ for $\Delta\neq[1,\dots,1]$. In next sections we prove, that this Hurwitz topological field theory generate the
Frobenius algebra that is the center $Z\left(\mathbb{C}[S_d] \right)$ of the group algebra $\mathbb{C}[S_d]$ of the
symmetric group $S_d$ and $U =\sum \limits_{\sigma\in S_d} \sigma^2$ \cite{AN2008}.

The definition of Hurwitz numbers can also be expanded onto a surface with a boundary in such a 
way that they generate Klein topological field theory \cite{AN, AN1, AN2008, AN3}. The relationship of
these new Hurwitz numbers with algebraic geometry and mathematical physics  is still not well understood
(some achievements in this direction are contained in \cite{N, KLN}, \cite{NO-LMP}).

\vspace {2ex}

\subsection {Hurwitz numbers and symmetric group.}

Describe now Hurwitz numbers $H_{e}^d(\Delta^1,\dots,\Delta^\f)$ in terms of the center $Z\mathbb{C}[S_d]$ of the
group algebra $\mathbb{C}[S_d]$ of the symmetric group $S_d$. The action of a permutation $\sigma\in S_d$ on a set
$T$ of $d$ elements splits $T$ into $\ell$ orbits consisting of $\Delta_1,\dots,\Delta_{\ell}$ elements, where
$\Delta_1+\dots+\Delta_{\ell}=d$. The Young diagram $[\Delta_1,\dots,\Delta_{\ell}]$ is called
\textit{cyclic type of $\sigma$}. All permutations of a cyclic type $\Delta$ form the   conjugate class
$C_\Delta\subset S_d$. Denote by $|C_\Delta|$ the number of elements in $C_\Delta$. The sum $\gC_\Delta$ of the
elements of the conjugate class $C_{\Delta}$ belongs to the center of the algebra
$Z\mathbb{C}[S_d]$. Moreover, the sums $\gC_\Delta$ generate the basis in the vector space $Z\mathbb{C}[S_d]$.

The correspondence $\Delta\leftrightarrow\gC_\Delta$ gives an isomorphism between the vector spaces $Y_d$ and
$Z\mathbb{C}[S_d]$. It transfers the structure of the algebra to $Y_d$. We will keep it in mind in this section,
speaking about multiplication on $Y_d$.

\vspace {1ex}

Describe now the Hurwitz number $H_{2}^d(\Delta^1,\dots,\Delta^\f)$ of the sphere $S^2$ in terms of the algebra $
Z\mathbb{C}[S_d]$. Consider different points $\{p_1, \dots, p_\f \}$ of $S^2$ and $ p\in S^2\setminus\{p_1, 
\dots, p_\f \}$. Consider the
standard generators of the fundamental group $\pi_1(S^2\setminus\{p_1,\dots, p_\f \},p)$. They are represented
by simple closed disjoint contours $\gamma_1,\dots,\gamma_\f$ 
which bypass respectively the points $p_1,\dots,p_\f $ and
each of which starts at $p$ and ends at $p$, and $\gamma_1\dots \gamma_\f=1$.

Consider now a covering $\varphi:\widetilde{\Omega}\rightarrow S^2$ of a given type $(\Delta^1,\dots,\Delta^\f)$ with
the critical values $p_1\dots p_\f$. The complete preimage of $\varphi^{-1}(p)$ consists of $d$ points $q_1,\dots,q_d$. 
The path along the contour $\gamma_i$ results in a permutation $\sigma_i\in S_d $ of $q_1,\dots,q_d$ . 
The conjugacy class of $\sigma_i$ is given by $\Delta^i$. Moreover, 
the product $\sigma_1\dots\sigma_\f$ is the identical
permutation. Thus, any covering of $S^2$ of type $(\Delta^1,\dots,\Delta^\f)$ generates an element of the set
$$
M=M(\Delta^1,\dots,\Delta^\f)=
\{(\sigma_1,\dots,\sigma_\f)\in(S_d)^\f|\sigma_i \in\Delta^i(i = 1,\dots,\f);\sigma_1\dots\sigma_\f=1\}.
$$
Moreover, equivalent coverings generate elements of $M$ that conjugated by some permutation $\sigma\in S_d$.

Construct now the inverse correspondence from the classes of $M(\Delta^1,\dots,\Delta^\f)$ to equivalent
classes of coverings $\varphi:\widetilde{\Omega}\rightarrow S^2$ of type $(\Delta^1,\dots, \Delta^\f)$ with the
critical values $p_1,\dots,p_\f$.
Cuts $r_i\subset S^2$ between points $p$ and $p_i$ inside the contour $\gamma_i$ generate a cut sphere
$\widehat{S}=S^2\setminus \bigcup\limits_{i=1}^d r_i$.

Correspond now the covering which corresponds to  $(\sigma_1,\dots,\sigma_\f)\in M$. For this we consider $d$
copies of the cut sphere $\widehat{S}$, number them, and glue its boundaries according to the permutations
$\sigma_1,\dots,\sigma_\f$. This gives a compact surface $P$. Moreover, the correspondances between the copies of
$\widehat{S}$ and $\widehat{S}$ generate the covering $\varphi:P\rightarrow S^2$, of type $(\Delta^1,\dots,\Delta^\f)$.
Conjugated by $\sigma\in S_d$ of the set $(\sigma_1,\dots,\sigma_\f)$ generate equivalent covering.

Thus
$$
H_{\e(S^2)}^d (\Delta^1, \dots, \Delta^\f) =
\sum_{\varphi\in\Phi(\Delta^1, \dots, \Delta^\f)} \frac {1} {|\texttt{Aut} (\varphi)|} =
\sum_{(\sigma_1, \dots,\sigma_\f) \in \widetilde{M}} \frac{1}{| \texttt {Aut} (\sigma_1, \dots,\sigma_\f)|}.
$$
where $\widetilde{M}$ is the set of the conjugated classes of $M$ by $S_d$ and $\texttt{Aut}(\sigma_1,\dots,\sigma_\f)$
is the stabilizer of $(\sigma_1, \dots,\sigma_\f)$ by these conjugations.

On the other hand,
$$
\sum_{(\sigma_1, \dots,\sigma_\f) \in \widetilde{M}} \frac{1}{| \texttt{Aut}
(\sigma_1, \dots,\sigma_\f)|}=\frac{1}{d!}| M (\sigma_1, \dots,\sigma_\f) |=<\Delta^1\dots\Delta^\f>.
$$

Thus,
\begin{equation}\label{sruct.cont} H_{\e(S^2)}^d (\Delta^1, \dots,\Delta^\f)= <\Delta^1\dots\Delta^\f>.
\end{equation}

For an  arbitrary closed connected surface $\Omega$ this relation turns into \cite{AN}
\begin{equation} H_{\e(\Omega)}^d (\Delta^1, \dots,\Delta^\f)= <\Delta^1\dots\Delta^\f U^{2-\e(\Omega)}>.
\end{equation}
where
$ U =\sum \limits_ {\sigma \in S_d}\sigma^2 $ \cite{AN, AN2008}.

The proof for arbitrary $ \Omega $ is almost the same as for the sphere.
You only need to replace the relation
$ \sigma_1 \dots \sigma_\f = 1 $ by the relations for the generators in $ \pi_1 (\Omega, p) $. 
For an orientable $ \Omega $, the new relation is $ [a_1, b_1] \dots [a_g, b_g] \sigma_1 \dots \sigma_\f = 1 $, while 
for a non-orientable $ \Omega $ the new relation has the form
$ c_1^2 \dots c_g^2 \sigma_1 \dots \sigma_\f = 1 $.

In particulary
\begin{equation}\label{HurGen}
H^d_{\e(\mathbb{R}P^2)} (\Delta^1, \dots,\Delta^\f)
= <\Delta^1\dots\Delta^\f U>.
\end{equation}

\vspace {2ex}

\subsection {Hurwitz numbers and representation theory. \label{Hurwitz numbers and representation theory}}

Formula (\ref{HurGen}) allows to describe Hurwitz numbers in terms of characters of the symmetric groups.
The corresponding formula is
\be\label{Mednykh}
H_{\e}^d (\Delta^1, \dots, \Delta^\f) =
(d!)^{-\e} |C_{\Delta^1}| \dots |C_{\Delta^\f}|\sum \limits _ {\chi} \frac { \chi (\gC_{\Delta^1})
\dots \chi (\gC_{\Delta^\f})} {\chi (1)^{\f-\e}}
\ee
where the summation ranges over all the characters of the irreducible representations of the group $ S_d $ and where
$ | {C}_\Delta | $ is the cardinality of the set of elements  of the cyclic type $ \Delta $.

The first versions of the formula in the language of symmetric groups appeared in the works of Frobenius and
Schur \cite{Fr, FS} (these were the cases later related to the unbranched coverings of $S^2$ and $\mathbb{RP}^2$). 
The geometric consideration for the case of arbitrary surfaces appeared in \cite{M1, M2}.
Now we give a sketch of the proof of the formula (\ref{Mednykh}).

Any partition $\lambda$ of weight $d$ generates the irreducible representation of $S_d$ of dimension $\dim \lambda$.
Let $\chi_\lambda$ be the character of this representation. Then $\dim\lambda =\chi_\lambda(\gC_{[1,\dots,1]})$.
For any Young diagram $\Delta$, we consider \textit{normalized character}:
\be\label{normalized-characters}
\varphi_\lambda(\Delta) :=| {C}_\Delta | \frac{\chi_\lambda(\Delta)}{\dim\lambda}.
\ee

The known orthogonality relations for the characters are \cite{Mac}
\be\label{orth1}
\sum_{\lambda} \left(\frac{{\rm\dim}\lambda}{d!}\right)^2\varphi_\lambda(\mu)\varphi_\lambda(\Delta) =
 \frac{ \delta_{\Delta,\mu} }{z_{\Delta}}
\ee
and
\be\label{orth2}
\left(\frac{{\rm\dim}\lambda}{d!}\right)^2
\sum_{\Delta} z_\Delta\varphi_\lambda(\Delta)\varphi_\mu(\Delta) =
\delta_{\lambda,\mu}
\ee
where $d=|\Delta|=|\lambda|$ and
\be\label{z_Delta}
z_\Delta=\prod_{i}m_i!i^{m_i} =\frac{d!}{|C_\Delta|}
\ee
is the order of the automorphism group of the Young diagram $\Delta$.
(In this formula $m_i$ is the number of rows of length $i$ in $\Delta$.)

Elements
\be\label{idempotents}
 \gF_\lambda = \left(\frac{{\rm\dim}\lambda}{d!} \right)^2\sum_\Delta z_\Delta\varphi_\lambda(\Delta) \gC_\Delta
\ee
form the basis of idempotents of $Z\mathbb{C}[S_d]$, that is
\be\label{idempotent}
\gF_\lambda \gF_\mu =0,\quad \mu\neq \lambda,\qquad
\gF_\lambda^2=\gF_\lambda
\ee
The transition between the two basises is given by
\be\label{class-via-idempotents}
 \quad \gC_\Delta = \sum_\lambda  \varphi_\lambda(\Delta)  \gF_\lambda
\ee
and therefore
\be
\gC_{\Delta^1}\cdot \gC_{\Delta^2}=\sum_\lambda \varphi_\lambda(\Delta^1)\varphi_\lambda(\Delta^2)\gF_\lambda
=\sum_\Delta H_2(\Delta^1,\Delta^2,\Delta)z_\Delta\gC_\Delta
\ee
Moreover
\be
< \gC_{\Delta^1}\cdots \gC_{\Delta^\f} U^{2-\e} > =\sum_\lambda
\varphi_\lambda(\Delta^1)\cdots \varphi_\lambda(\Delta^\f)< \gF_\lambda U^{2-\e} >
\ee
and
\be
< \gF_\lambda U^{2-\e}> =\left( \frac{{\rm dim}\lambda}{|\lambda|!}\right)^{\e}
\ee

Therefore
\[
H_{\e({\Sigma})}(\Delta^1, \dots, \Delta^\f) = < \gC_{\Delta^1}\cdots \gC_{\Delta^\f}>_{{\Sigma}} =
< \gC_{\Delta^1}\cdots \gC_{\Delta^\f} U^{2-\e({\Sigma})} >
\]
\be\label{Mednykh-Hurwitz}
=\sum_\lambda
\varphi_\lambda(\Delta^1)\cdots \varphi_\lambda(\Delta^\f)\left( \frac{{\rm dim}\lambda}{|\lambda|!}\right)^{\e}
\ee
that is equivalent to (\ref{Mednykh}).

From (\ref{Mednykh-Hurwitz}) and (\ref{orth1}) we get
$$
 H_2(\Delta^1,\Delta) =\frac{\delta_{\Delta^1,\Delta}}{z_\Delta}
$$
and for symmetric group $A=S_d$ and vectors $\Delta^1,\Delta$ we obtain
\be
F_A^{\Delta^1,\Delta} = z_\Delta \delta_{\Delta^1,\Delta}
\ee

This formula has a simple geometric explanation in the framework of Hurwitz topological field theory. 
Quadratic
form $F_A^{\Delta,\Delta}$ is used in the cutting axiom for gluing two parts of a surface.
It is diagonal form $kI$, where $k$ is inverse to the number of all admissible ways of gluing between the boundaries of
the two parts. 
Moreover, gluing along a contour on which the covering has the same degree is admissible.
Therefore $k=z_\Delta$ from (\ref{z_Delta}). According to our definitions,  Hurwitz topological field theory
generates the Frobenius algebra with the structure constants
$\sum\limits_\Delta H_2(\Delta^1,\Delta^2,\Delta) F_A^{\Delta,\Delta^3}$ in the basis of Young diagrams $\{\Delta\}$.
According to (\ref{sruct.cont}), these structure constants coincide with structure constants of $Z(\mathbb{C}[S_d])$
in basis $\gC_{\Delta}$. The correspondence $\Delta\leftrightarrow\gC_{\Delta}$ generates the isomorphism between
the Frobenius algebra of Hurwitz topological field theory and the centre of group algebra of symmetric group.
(see\cite{AN} for details).

For the number $D(\Delta)$ describing M\"{o}bius cut (\ref{Moebius-cut}), we get
\be\label{D(Delta)}
D(\Delta)=z_\Delta H_1(\Delta)
\ee
where $H_1(\Delta)$ is the Hurwitz number counting the coverings of the real projective plane
$\mathbb{RP}^2$ with one critical value with the ramification profile $\Delta$.

Formulas (\ref{Mednykh-Hurwitz}) and (\ref{orth2}) allow us to give an independent proof of the fact
that Hurwitz numbers satisfy the axioms of
topological field theory:

\noindent
\bp\label{Hurwitz-down-Lemma}
Let us define numbers $H_{\e({\Sigma})}(\Delta^1, \dots, \Delta^\f)$ by
(\ref{Mednykh-Hurwitz}).
Consider the set of partitions $\Delta^i,\,i=1,\dots,\f_1+\f_2$ of the same weight $d$. We have 
the handle cut relation (Fig \ref{f3}):
\begin{eqnarray}\label{handle-cut}
H_{\e  -2}(\Delta^{1},\dots,\Delta^{\f}) = \sum_{\Delta\atop |\Delta|=d}
H_{\e}(\Delta^{1},\dots,\Delta^{\f},\Delta,\Delta)z_\Delta
\\
= \sum_{\Delta\atop |\Delta|=d}\frac{
H_{\e}(\Delta^{1},\dots,\Delta^{{\f}},\Delta,\Delta)}
{H_2(\Delta,\Delta)}\,.\nonumber
\end{eqnarray}
(that is the manifistation of (\ref{cut}) ),  and we have surface cut relation (Fig \ref{f2}):
\begin{eqnarray}\label{Hurwitz=Hurwirz-Hurwitz}
H_{\e_1+\e_2 -2}(\Delta^{1},\dots,\Delta^{{\f}_1+{\f}_2}) = \sum_{\Delta\atop |\Delta|=d}
H_{\e_1}(\Delta^{1},\dots,\Delta^{{\f_1}},\Delta)z_\Delta
H_{\e_2}(\Delta,\Delta^{{\f}_1+1},
\dots,\Delta^{{\f}_1+{\f}_2})\\
 = \sum_{\Delta\atop |\Delta|=d}\frac{
H_{\e_1}
\left(
\Delta^{1},\dots,\Delta^{\f_1},\Delta
\right)
H_{\e_2}
\left(
\Delta,\Delta^{\f_1+1},\dots,\Delta^{\f_1+\f_2}
\right)
}
{H_2(\Delta,\Delta)}\,.\nonumber
\end{eqnarray}
(that is can be either cuts given by (\ref{cut})  and (\ref{Klein-cut})), or the Moebius 
cut (\ref{Moebius-cut}).
\begin{eqnarray}
\label{Hurwitz-down}
H_{\e-1}(\Delta^{1},\dots,\Delta^{{\f}})=
\sum_{\Delta}\,
H_{\e}(\Delta^{1},\dots,\Delta^{{\f}},\Delta) D(\Delta)\\
=
\sum_{\Delta}\,\frac{
H_{\e}(\Delta^{1},\dots,\Delta^{{\f}},\Delta)  H_{1}(\Delta)}{H_2(\Delta,\Delta)}\quad ,\nonumber
\end{eqnarray}
where $\frac {H_{1}(\Delta)}{H_2(\Delta,\Delta)}= D(\Delta) $ 
are rational numbers:
\be\label{delta(Delta)}
D(\Delta)= z_\Delta H_{1}(\Delta)
=\sum_{\lambda \atop |\lambda|=|\Delta|} \chi_\lambda(\gC_{\Delta})
\ee
see (\ref{Mednykh}).
\ep

\noindent For instance, we get that the Hurwitz numbers of the projective plane can be obtained from
the Hurwitz numbers of the Riemann sphere, while the Hurwitz numbers of the torus and  the Klein bottle
can be obtained from the Hurwitz numbers of the projective plane.

\paragraph{Jucys-Murphy elements.} Jucys-Murphy elements serves to relate Hurwitz numbers
to classical and also to the simplest quantum integrable systems (see Subsection \ref{Dubr-quant}  below).
First applications of Jucys-Murphy elements to the description of integrable systems (namely, to the 
so-called KP and to the so-called TL hierarchies of integrable equations) were found
in \cite{Goulden-Paquet-Novak}, \cite{GayPakettHarnad1} and in most clarified way in \cite{HarnadOverview}.

Let us consider the sums of transpositions as follows:
\[
\gJ_{m} = (1,m) + (2,m) +\cdots + (m-1,m),\quad m=2,\dots, d
\]
(one implies $\gJ_1 =0$)
which are known as Jucys-Murphy elements of the group algebra $\mathbb{C}[S_d]$ introduced 
in \cite{Jucys},\cite{Murphy}.
Jucys-Murphy elements do not belong to $Z \mathbb{C}[S_d]$, however they
pairwise commute and generate the maximal abelian subalgebra of $\mathbb{C}[S_d]$
(Gelfand-Tseitlin algebra). Moreover \cite{OkounkovVershik}, any symmetric function
of $\gJ_1,\dots,\gJ_d$  belongs to $Z\mathbb{C}[S_d]$ and
\be\label{Jucys-Murphy-content}
G(\gJ_1,\dots,\gJ_{d}) \gF_\lambda = G(c_1,\dots,c_d) \gF_\lambda
\ee
where $G$ is a symmetric function of the arguments, and
$c_1,\dots,c_d$ is the set of the {\it contents} of all $d$ nodes of $\lambda$.
(The content of the node of Young diagram $\lambda$ with coordinates $(i,j)$ is
defined as $j-i$\cite{Mac}).
In other words, we get
\be\label{J-via-F}
 G(\gJ_1,\dots,\gJ_d)= \sum_\lambda G(c_1(\lambda),\dots,c_d(\lambda))\gF_\lambda
 =\sum_\Delta G^*(\Delta)\gC_\Delta
\ee
where
\[
 G^*(\Delta) = \sum_\lambda \left(\frac{{\rm dim}\lambda}{d!} \right)^2
 \varphi_\lambda(\Delta)
 G(c_1(\lambda),\dots,c_d(\lambda))
\]

We are going to explain the following:

A) Integrable hierarchies generate correlators of the following type:

The Kadomtsev-Petviashvili (KP) hierarchy \cite{NovikovManakovZakharov}
generates (see Proposition \ref{Prop-Jucys-Murphy-tau-TL})
\be\label{G-C-KP}
  <G(\gJ_1,\dots,\gJ_d)\gC_{\Delta^1}>_{S^2}
\ee
 The two-component version of the KP hierarchy (and the relativistic Toda lattice hierarchy) generates
  \be\label{G-C-TL}
  <G(\gJ_1,\dots,\gJ_d)\gC_{\Delta^1}\gC_{\Delta^2}>_{S^2}
\ee
The BKP hierarchy \cite{KvdLbispec} generates (see Proposition \ref{Prop-Jucys-Murphy-tau-BKP})
\be\label{BKP}
  <G(\gJ_1,\dots ,\gJ_d)\gC_{\Delta^1}>_{\mathbb{R}R^2}
\ee
Here $G$ is any (symmetric) function defined by the choice of the solution of the equations
of the hierarchies. 

B) The insertion of $G$ can be equivalently described with help of the so-called {\it completed cycles}. This
link will be discussed (see Lemma \ref{JMcontent-q}).

C) The insertion of $G$ can be obtained as a result of the action of certain vertex operators on generating
functions for Hurwitz numbers. Sometimes (see Subsection \ref{Dubrovin2}) the generating function can be 
interpreted as the so-called tau 
function of the integrable hierarchy. In this case this action is known
as the action of the additional symmetries of classical integrable systems studied in \cite{Orl1987}.
Then the vertex operators can also be considerd  treated as evolutionary operators for certain simple 
quantum models, see Subsection \ref{Dubr-quant}.

Finally, at the end of this subsection, we would like to mention the excellent book \cite{ZL}, 
which discusses many related topics.

\subsection {The generating function for Hurwitz numbers.}
\vspace {2ex}

Important applications of the Hurwitz numbers are related to the corresponding generating functions for 1- and 2-
Hurwitz numbers. The (disconnected) 1-Hurwitz number $h_ {m, \Delta}^{\circ}$ is the notation for the Hurwitz number
$ H_2^d (\Delta, \g_1, \dots, \g_m) $, where $\g_1=\dots=\g_m=[2,1, \dots, 1]$.

The generating function for the 1-Hurwitz numbers depends on the infinite number of formal variables $ p_1, p_2, \dots $.
We associate a Young diagram  $ \Delta $ with rows of length $ \Delta_1, \dots, \Delta_k $ with the monomial 
$ p_{\Delta} = p_ {\Delta_1}, \dots, p_ {\Delta_k} $. The generating function for the 1-Hurwitz numbers is defined as
$$ 
F^{\circ} (u | p_1, p_2, \dots) =\sum \limits_{m = 0}^\infty
\sum \limits_{\Delta}^\infty \frac {u^m} {m!} h_ {m, \Delta}^{\circ} p _ {\Delta}. 
$$
This function has a number of remarkable properties that have been discovered relatively recently.
The first one is the relationship
\begin {equation}\label{K-a-J}
\frac{\partial F^{\circ}} {\partial u} = L^{\circ}F^{\circ},
\end{equation}
where
\be\label{Hopf}
L^{\circ} = \frac {1} {2}\sum \limits_ {a, b = 1}^\infty \Big ((a + b) p_ap_b
\frac {\partial} {\partial p_ {a + b} } + ab p_ {a + b} \frac {\partial^2} {\partial p_a \partial p_b} \Big).
\ee
This relationship was first discovered in \cite{GJ} by purely combinatorial methods. But it also has a geometric
explanation \cite{MM1}, \cite{MM3}.
Consider a covering $\varphi:{\Sigma}\rightarrow S^2$ of the type
$(\Delta, \g_1, \dots, \g_m)$. Let $q, p\in S^2$ be the critical points of the covering $\varphi$ related 
to the Young diagrams $\Delta$ and $\g_m$, respectively. Connect the points $q$ and $p$ with a line $l$ without 
self-intersections. The preimage of $\varphi^{-1}(l)$ consists of $d-1$ connected components, exactly one of which, say 
$\tilde{l}$, contains the critical point $\tilde {p}$ with the critical value $p$. The ends of the component $
\tilde {l}$ are the pre-images $\tilde {q}_1$ and $\tilde {q}_2$  of $q$.

Now we move the point $ p $ along the line $ l $ in the direction of the point $ q $, respectively, 
continuously changing the covering $ \varphi $.
 As a result, we get a covering $ \varphi '$ of the type
$ (\Delta', \g_1, \dots, \g_ {m-1}) $. 
Let's see which Young diagram $ \Delta '$ can serve this.
Let $ \tilde {q}_1 = \tilde {q}_2 $ and $ c $ be the branching order of the covering $ \varphi $ at this point $
\tilde {q} = \tilde {q}_1 = \tilde { q}_2 $. 
The orders of the critical points other than $ \tilde {q} $,
will not change if the cover $ \varphi $ is deformed into a cover $ \varphi '$.
As a result of deformation,
the point $ \tilde {q} $ is splitted into 2 points with branch orders $ a $ and $ b $,
where $ a + b = c $. Thus, the monomial $ p_ \Delta $ becomes a monomial
$ p_ap_b \frac {\partial p_ \Delta} {\partial p_c} $.

Suppose that the critical points $ \tilde {q}_1 $ and $ \tilde {q}_2 $ do not coincide and the orders of their
branching are $ a $ and $ b $, respectively. Then, as before, in the process of deformation of covering 
$ \varphi $ into a covering $ \varphi '$, the orders of critical points other than $ \tilde{q}_1 $ and $
\tilde {q}_2 $ will not change. As a result of the deformation, the points $ \tilde {q}_1 $ and $ \tilde {q}_2 $ 
will be transferred to one critical point of order $ c = a + b $. Thus, the monomial $ p_ \Delta $ becomes a monomial
$ p_c\frac {\partial^2 p_ \Delta} {\partial p_ap_b}$.
Summation over all possible equivalence classes of the covers with the profile $ (\Delta, \g_1, \dots, \g_m) $
and all their deformations into covers with all possible profiles $ (\Delta ', \g_1, \dots, \g_ {m-1} ) $ results in
the relation (\ref{K-a-J}).

Differential properties the function $F^{\circ} (u | p_1, p_2, \dots)$ investigated in 
\cite{MM1,MM3,MM4,MM5,AMMN-2011,AMMN-2014}.

\vspace{2ex}

We defined the Hurwitz numbers, as the weighted number (\ref{disconH}) of coverings $\varphi:P\rightarrow{\Sigma}$
where the surface $P$ is not necessarily connected. Such Hurwitz numbers are often called 
\textit{disconnected Hurwitz numbers}. If in this definition we consider only connected surfaces $P$, 
then the resulting number is called  \textit{connected Hurwitz number}.
Let us denote $h_ {m, \Delta}^{\bullet}$  the connected version of 1-Hurwitz number $h_ {m, \Delta}^{\circ}$.
It was proved that its generating function
$$ F^{\bullet} (u | p_1, p_2, \dots) =\sum \limits_{m = 0}^\infty
\sum \limits_{\Delta}^\infty \frac {u^m} {m!} h_ {m, \Delta}^{\bullet} p _ {\Delta}$$
is a solution to the KP hierarchy. By simple combinatorial methods it is easy to prove that
$$F^{\bullet} (u | p_1, p_2, \dots)=\ln(F^{\circ} (u | p_1, p_2, \dots)).$$
Therefore (\ref{K-a-J}) generates
\begin {equation}\label{conn-K-a-J}
\frac{\partial F^{\bullet}} {\partial u} = 
\frac {1} {2}\sum \limits_ {a, b = 1}^\infty 
\Big ((a + b) p_ap_b\frac {\partial F^{\bullet}} {\partial p_ {a + b} } +
abp_{a+b} \frac{\partial F^{\bullet}}{\partial p_a}
\frac{\partial F^{\bullet}} {\partial p_b} + 
abp_{a + b} \frac {\partial^2 F^{\bullet}} {\partial p_a \partial p_b} \Big).
\end{equation}

\subsection{Gaussian integrals over sets of complex matrices \label{random-complex}}

On this subject there is an extensive literature, for instance see \cite{Ak1,Ak2,AkStrahov,S1,S2}.

We will consider integrals over $N\times N$ complex matrices $Z_1,\dots,Z_n$ where the measure is defined as
\be\label{CGEns-measure}
d\Omega(Z_1,\dots,Z_n)= c_N^n 
\prod_{i=1}^n\prod_{a,b=1}^N d\Re (Z_i)_{ab}d\Im (Z_i)_{ab}\text{e}^{-N|(Z_i)_{ab}|^2}
\ee
where the integration domain is $\mathbb{C}^{N^2}\times \cdots \times\mathbb{C}^{N^2}$ and where $c_N^n$
is the normalization 
constant defined via $\int d \Omega(Z_1,\dots,Z_n)=1$.

The set of $n$ $N\times N$ complex matrices
and the measure (\ref{CGEns-measure}) is called {\it $n$ independent complex 
Ginibre ensembles}, such ensembles have wide applications in physics  and in 
information transfer theory \cite{Ak1},\cite{Ak2},\cite{AkStrahov}
\cite{S1},\cite{S2},\cite{Alfano}.

\paragraph{The Wick rule.}
Let us recall the following property of the Gaussian integrals: 

From
\[
 c\int_{\mathbb{C}} z^d{\bar z}^m e^{-N|z|^2} d \Im z d\Re z = d! \delta_{d,m} N^{-d}
\] 
($c$ is the normalization constant)
it follows that $ d! $ that enters the right hand side can be interpreted as the number of ways to split 
the product $ z \cdots z \bar {z} \cdots \bar{z} $ into the pairwise products $ z  \bar{z} $. 
 This funny observation is very useful and is known in physics 
as the Wick rule. The Wick rule is associated with the interpretations of matrix models in many problems of physics, 
for example, in the papers on statistical physics \cite{Itzykson-Zuber}
quantum gravity \cite{BrezinKazakov} and also in \cite{Chekhov-2014},
\cite{ChekhovAmbjorn} which precede our work.

 In what further, it will be applied to the product of matrix entries $\left( Z_i \right)_{a,b}$ and its
 complex conjugate
 $\left( \bar{Z}_i \right)_{a,b}$, $i=1,\dots,n$, $a,b=1,\dots,N$.
Then, the thanks to the measure $d\Omega$, the integral of a monomial in matrix entries either vanishes,
or is equal to a power of $N$.

The integrands we will consider  in Sections \ref{Geom comb Hurwitz},\ref{Hurwitz-quantum-models} are polynomials
in entries of matrices $(Z_iA_i)$ and $(\hat{Z}_i\hat{A}_i)$, $i=1,\dots, n$
(in certain cases integrands can be formal series in polynomials)
where matrices $A_i$ and $\hat{A}_i$ are not necessarily related while each $\hat{Z}_i$ is 
Hermitian conjugate $Z_i$.

\section{Integrals and Hurwitz numbers \label{Integrals-Hurwitz} }

\subsection{Geometrical and combinatorial definition of Hurwitz numbers via graphs \label{Geom comb Hurwitz}}

Let $ \Sigma $ be a connected compact orientable surface without boundary of the Euler characteristic $ \e $.
We fix on $\Sigma$ two nonempty sets of points. The points of the first set $ \{c^1, \dots, c^\f \} $ will be called 
\textit{capitals}, and the points of the second set $ \{w_1, \dots, w_\V \} $ will be called \textit{watch towers}. 
To each capital $ c^j $ we assign a Young diagram $ \Delta^j = [\Delta^j_1, \dots, \Delta^j_{\ell (j)}] $ with rows
$ \Delta^j_1 \geq \Delta^ j_2, \dots, \geq \Delta^j_{\ell (j)}> 0 $ from the set $ \Upsilon_d $ of all Young 
diagrams of weight $ d $.

Consider a graph $ \Gamma $ with vertices $ \{w_1, \dots, w_\V \} $ on $ \Sigma $. We require that 
the edges of the graph do not intersect  at interior points, and that the complement to the edges 
disintegrates 
into connected, simply connected domains $ P^j \ni c^j (j = 1, \dots, \f) $. (Such a partition always exists, 
except when $ \Sigma $ is a sphere and $ \f = \V = 1 $.) 
We denote the number of edges of $\Gamma$ by $n$, the Euler characteristic of $\Sigma$ is $\e=\f+\V-n$.
We will call the domains $ P^j $ \textit{basic polygons
}.
The boundary of each basic polygon consists of \textit{sides} generated by the edges of the graph and \textit{vertices} 
generated by the vertices of the graph.
Thus, an edge of the graph generates either two different side of the same basic polygon or two sides of two different 
basic polygons.
A vertex $ w $ of the graph (the watchtower)
generates $ k $ vertices of basic polygons, where $ k $ is the number of basic polygons with vertex $ w $.
Fix one of vertices $ v $ of each basic polygon $ P $.

\vspace {0.1cm}

Consider a  $ d $ -sheeted branched covering of the basic polygon $ P^j $, corresponding to the Young diagram $ \Delta^j $
having a single critical value $ c_j $. This covering consists of $ \ell (j) $ $ \Delta^j_i $ -sheeted 
cyclic covers $ \varphi^j_i: P^j_i \rightarrow  P^j $. The preimages of the sides of the basic polygon $ P^j $ are 
called sides of the polygon $ P^j_i $.
The order $ | \texttt{Aut} (\varphi^j) | $ of the automorphism group of the covering
$ \texttt{Aut} (\varphi^j) $ is equal to the order of the automorphism group of the Young diagram $
\Delta^j $ i.e.
$ | \texttt{Aut} (\Delta^j) | = m^j(1)! \dots m^j (\Delta^j_1)! \Delta^j_1 \Delta^j_2, \dots,
\Delta^j_ {\ell (j)} $,
where $ m^j (r) $ is the number of rows of length $ r $ in the Young diagram $ \Delta^j $.

We divide the sides of the set of polygons $ \{P^j_i\} $ into pairs so that the images of the both sides of a pair 
coincide under the action of the maps $ \varphi^j_i $ and belong to the closure of only one basic polygon 
$ P^j $ if and only if each side of the pair does not belong to the polygon $ P^k_i $ where $ k \neq j $. 
Glue the sides of one pair so that the images of glued points coincide under the action of the covers 
$ \varphi^j_i $. We call such gluing systems
\textit{admissible}. The number of different ways of gluing of the preimages of an edge of $\Gamma$ is $ d! $. 
Therefore,  the total number of admissible gluing systems equals $ (d!)^n $.

Each admissible gluing system $ \xi $ generates a branched covering
$ \varphi (\xi): \Sigma (\xi) \rightarrow{\Sigma} $. The surface $ \Sigma (\xi)$ that is glued  from 
polygons $ P^j_i $   is compact, orientable, but possibly not connected. Critical values of covering
$ \varphi (\xi) $ lie in the set $ \{c^1, \dots, c^{\f} \} \cup \{w_1, \dots, w_{\V} \} $, and the topological 
type of the critical value $ c^j $ is equal to $ \Delta^j $. Topological type $ \widetilde{\Delta}_k $
of critical values $ w_k $ depends on the gluing system.
Automorphisms of the coverings $ \varphi^j $ transfer admissible gluing into admissible ones, 
preserving the topological type of the glued covers.

Now consider an arbitrary covering $\varphi: \hat{\Sigma} \rightarrow \Sigma $, the complement
$ \Sigma^0 = \Sigma \setminus \Gamma $ and the preimage $ \hat {\Sigma}^0 = \varphi^{- 1}(\Sigma^0) $. Adding
the preimage  $ \varphi^{- 1} (\Gamma) $ implements the admissible gluing system of the surface 
$ \hat{\Sigma}^0 $ to get the surface $ \hat{\Sigma} $. 
Thus, an admissible gluing system allows you to get any branched cover from any
equivalence class of coverings of the surface $\Sigma$ with critical values 
$ \{c^1, \dots, c^{\f} \} \cup \{w_1, \dots, w_{\V} \} $ of topological type $ \Delta^j $ at points $ c^j $.
Denote by $ \Phi^{\V} (\Delta^1, \dots, \Delta^{\f}) $ the set of equivalence classes of all such coverings.

The correspondence $ \xi \mapsto \varphi (\xi) $ generates a mapping 
$ \Psi $ of the set $ \Xi (\Delta^1, \dots, \Delta^\f) $ of
all admissible gluing of the sides of the polygons $ P^j_i $ on the set $ \Phi^\V(\Delta^1, \dots, \Delta^\f) $.
The mapping $ \Psi $ is constant on the orbits of the action of the group $ \texttt{Aut} = 
\prod_{j} \texttt{Aut} (\Delta^j) $.
In addition, the kernel of the action of the group $ \texttt{Aut} $ on the set $ \Xi (\Delta^1,
\dots, \Delta^\f) $ coincides with
automorphisms of the coverings $ \varphi(\xi) $. Thus,

$$
\frac{(d!)^n}{\prod_{j}z_{\Delta^j}}= \frac{|\Xi(\Delta^1,\dots,\Delta^\f)|}{|\texttt{Aut}|}=
\sum\limits_{\varphi\in \Phi^\V(\Delta^1,\dots, \Delta^\f)} \frac{1}{|\texttt{Aut}(\varphi)|}=
\sum\limits_{\widetilde{\Delta}^1,\dots,\widetilde{\Delta}^\V\in \Upsilon_d}
H_{\Sigma}(\Delta^1,\dots, \Delta^\f, \widetilde{\Delta}^1,\dots, \widetilde{\Delta}^\V).
$$

\vspace{0.3cm}

In order to find a specific Hurwitz number $ H_{\Sigma} (\Delta^1, \dots, \Delta^\f,
\widetilde {\Delta}^1, \dots, \widetilde{\Delta}^\V) $ we assign $ N \times N $ matrices  to the sides and 
corners of the basic polygons.

The matrix assigned to a side $ l $ of a basic polygon $ P $ is denoted by $Z^{l, P} $.
Thus, the edge $ l $ separating the basic polygons $ P', P''$ corresponds to the matrices $ Z^{l, P'} $ 
and $ Z^{l, P''} $. The case $ P'= P''$ requires a separate discussion.
In this case, the matrices $ Z^{l, P'} $ and $ Z^{l, P''} $ correspond to the sides of the same basic polygon 
and these sides are identified on the surface $ \Sigma $. In all cases, we choose the matrices $ Z^{l, P'} $
and $ Z^{l, P''} $ Hermitian conjugate. We get $n$ pairs of Hermitian conjugate complex matrices, this
collection we denote $\{Z\}$.

At the vertex of the basic polygon $ P $ is the watchtower $ w $. Denote by $ A^{w, P} $ the matrix 
assigned to this vertex of the basic polygon. We will call these matrices \textit{source matrices}. One gets
a set of
matrices $ A^{w, P_1}, \dots, A^{w, P_k} $ related to vertex $w$, where $ P_1, \dots, P_k $ are
basic polygons adjacent to the watchtower $ w $.
The collection of $2n$ source matrices we denote $\{ A\}$.

The matrices $ Z^{l_1, P}, \dots, Z^{l_s, P} $ and $ A^{w^1, P}, \dots, 
A^{w^s, P} $ correspond to the basic polygon $P$ with the capital $c$.
We assume that the indexes of the sides and vertices are ordered counterclockwise, with the vertex $ w^i $ lying between
sides $ l_i $ and $ l_{i + 1} $. Denote by $ M(\{ZA\}^P) = Z^{l_1, P} A^{w^1, P}
 \dots Z^{l_s, P} A^{w^s, P} $ the product of the matrices corresponding to polygon border bypass, 
 we call this product \textit{monodromy around capital} $c$.
 Here $ \{ZA \}^P $ denotes the set of all pairwise products $ Z^{l_i, P} A^{w^i, P} $, 
 where the matrix $ Z^{l_i, P} $ assigned to the side $ l_i $ belonging to the country $ P $, 
 and the matrix $ A^{w^i, P} $ is assigned to the vertex $ w_i $, which is the endpoint  of the side 
 $ l_i $ (remember that the sides are oriented).
  In what follows we will write $ M (c)$ instead of $ M(\{ZA\}^P)$ where $c$ is the capital
of the polygon $P$.

 Consider the Young diagram $ \Delta = [\Delta_1, \dots, \Delta_ {\ell}] $. The automorphism group 
 $ \texttt{Aut} (\Delta) $ has the order $ | \texttt{Aut} (\Delta) | = \prod \limits_{i = 1}^{\ell}
 \Delta_i \prod \limits_{j = 1}^{d} k_j! $, where $ k_j $ is the number of  rows of a Young diagram of length $ j $. 
 To a capital $c_i$ and to the related Young diagram $ \Delta^i = [\Delta_1^i, \dots, 
\Delta_{\ell}^i] $ we assign the following polynomial in matrix entries 
$$
M_{c_i}^{\Delta^i}=\frac{1}{|\texttt{Aut}(\Delta)|} \texttt{tr}((M(c_i))^{\Delta_1^i})\dots 
\texttt{tr}((M(c_i))^{\Delta_\ell^i}),
$$
Below $\{ZA\}$ is the collection of all $\{ZA\}^P$ related to the set of all polygons $P_1,\dots,P_\f$.
Put
\be\label{G}
M(\Gamma,\{ZA\},\{\Delta\})=\prod\limits_{i=1}^\f  M_{c_i}^{\Delta^{i}}
\ee
Each edge of the graph corresponds to a pair of Hermitian conjugate matrices. Their union belongs to the set
$ \mathcal {Z} $, consisting of $ n $ pairs of Hermitian conjugate matrices, where $ n $ is the number of 
edges of the graph $ \Gamma $.
The right side of the formula (\ref{G}) contains $ d $ sets of matrices $ \{Z \} $. Moreover, to each edge 
$ l $ of the graph $ \Gamma $ corresponds to $ d $ pairs of Hermitian conjugate matrices 
 $(Z^{l,1},\hat{Z}^{l,1},\dots,Z^{l,d},\hat{Z}^{l,d}) \in\mathcal{Z}^{d}$.
 We denote their matrix elements
 $Z^{l,i}_{\alpha\beta},\hat{Z}^{l,i}_{\alpha\beta}$.
 Put $U^{l;s,t}_{\alpha,\beta}=Z^{l,s}_{\alpha\beta}\hat{Z}^{l,t}_{\beta\alpha}$.

Now on the set $ \mathcal{Z}^{d} $ we define the measure
 $ d \Omega $, assuming that the integral of a monomial product of
entries of matrices from  $ \mathcal{Z} $ is equal to $N^{-d}$ if it is a product of monomials of 
the form $ U^{l; s, t}_{\alpha,\beta} $ with respect to
all edges of the graph $ \Gamma $. The integral of the remaining monomials is assumed to be 0.
(This measure is exactly the mesaure introduced in Subsection \ref{random-complex}).
Put
$$
I(\Gamma,\{A\},\{\Delta\})=
\int\limits_{\mathcal{Z}^{d}} M(\Gamma,\{ZA\},\{\Delta\})d\Omega.
$$
where $\{ A\}$ is the collection of all source matrices and where $\{\Delta\}$ is the set of Young diagrams
$\Delta^1,\dots,\Delta^\f$ of the weight $d$.

\vspace {0.1cm}

We now consider an arbitrary watchtower $ w_v$  and the matrices 
$ A^{w_v, P_1}, \dots, A^{w_v, P_k} $ associated with it.
We assume that the indexes of the basic polygons $ P_i $ correspond to the clockwise rounds around  
the watchtower $ w_v $.
Denote by $ {\cal{A}}_{w_v} = A^{w_v, P_1} \dots A^{w_v, P_k} $ the matrix product.
We will call this product \textit{monodromy around the watchtower} $w_v$.

We associate the Young diagram $ \Delta = [\Delta_1, \dots, 
\Delta_{\ell}] $ with a polynomial of products of matrix elements
$$
 {\cal{A}}_{w_v}^{\Delta}= \texttt{tr}({\cal{A}}_{w_v}^{\Delta_1})\dots \texttt{tr}({\cal{A}}_{w_v}^{\Delta_\ell})
$$

\begin{Theorem}\label{Theorem}
$$
I(\Gamma,\{A\},\{\Delta\})=N^{-nd}
\sum\limits_{\widetilde{\Delta}^1,\dots,\widetilde{\Delta}^\V\in \Upsilon_d}{\cal{A}}_{w_1}^{\widetilde{\Delta}^1}
\dots {\cal{A}}_{w_\V}^{\widetilde{\Delta}^\V}
H_{\Sigma}(\Delta^1,\dots, \Delta^\f, \widetilde{\Delta}^1,\dots, \widetilde{\Delta}^\V).
$$
\end{Theorem}

As we can see, $ I(\Gamma, \{ A \}, \{ \Delta \}) $ depends on the eigenvalues of special products of 
the original matrices, namely, on the eigenvalues of the watchtower monodromies 
$ {\cal {A}}_{w_1}, \dots, {\cal {A}}_{w_\V} $.

\begin{proof} 
 If the integral of the monomial of matrix elements is not equal to 0, then its part,
generated by the matrices $ Z $ and $ \hat{Z} $ is generated by monomials
$U^{l;s,t}_{\alpha}=Z^{l,s}_{\alpha\beta}\hat{Z}^{l,t}_{\beta\alpha}$.
These monomials disappear after
integration and contribute giving the factor $N^{-nd}$. Grouping the remainder of the monomial by watchtowers, 
we obtain a monomial polynomial of the form
\begin{equation}\label{I}
\prod\limits_{i=1}^V  {\cal{A}}_{w_i}^{\widetilde{\Delta}^i},
\end{equation}
where
$$  {\cal{A}}_{w_v}^{\Delta}= \texttt{tr}(  {\cal{A}}_{w_v}^{\Delta_1})\dots 
\texttt{tr}( {\cal{A}}_{w_v}^{\Delta_\ell})$$
for the Young diagram $\Delta=[\Delta_1,\dots, \Delta_{\ell}]$.

We now describe the expression (\ref{I}) in terms of the coverings of the surface $ \Sigma $.
Consider the set of $ \{l \} $ edges of the graph $ \Gamma $ and its complement $ \Sigma^0 = \Sigma \setminus \{l \} $.
It splits into basic polygons. Consider the covering $ \hat {\Sigma^0}
 \rightarrow \Sigma^0 $ with the critical value  in the capitals $ \{c^j \} $ of basic polygons of topological type
 $ \{\Delta^j \} $. The integration of the monomial
$ U^{l; s, t}_{\alpha,\beta} $ glues together the sides $ s $ and $ t $ covering the edge $ l $. Thus, 
the integral of the monomial generates a covering $ \hat{\Sigma} \rightarrow \Sigma $ of degree $ d $ 
and a monomial of a polynomial of the form (\ref {I}).
Denote by $ \mathcal {K} $ the set of coverings constructed in this way.

Let $ \Delta $ be the Young diagram of a basic polygon $ P $. Consider the restriction $ \hat{P} \rightarrow P $ of the
coverings of $ \hat {\Sigma^0} \rightarrow \Sigma^0 $ on the  preimage of the basic polygon $ P $. 
Group $ \texttt{Aut} (\Delta) $ acts on the cover of $ \hat {P} \rightarrow P $ by automorphisms. 
This action changes the gluing system and therefore
acts on the set $ \mathcal {K} $, preserving the equivalence class of the covering. Denote by $ \Phi $ the set
of equivalence classes of coverings from $ \mathcal{K} $.

Suppose that all covers $ \varphi \in \mathcal {K} $ have the trivial automorphism group.
Then the set $ \mathcal {K} $ splits into
\begin{equation}\label{H}
\frac{|\mathcal{K}|}{\prod\limits_{i=1}^\f|\texttt{Aut}(\Delta_i)|}
\end{equation}
equivalence classes. The automorphism of the covering $ \varphi $ is realized by replacing the gluing system 
preserving the equivalence class of the covering. Thus, in the general case, the formula (\ref{H}) gives the sum
$$
\sum\limits_{\varphi\in\Phi}\frac{1}{|\texttt{Aut}(\varphi)|}.
$$

Summation over all monomials gives the statement of the theorem.

\end{proof}

\begin{Corollary}
If the size of the matrices $ N $ is large enough, then the integral 
$ I (\Gamma, \{A \}, \{\Delta\}) $ allows you to find all 
Hurwitz numbers
$H_{\Sigma}(\Delta^1,\dots, \Delta^\f, \widetilde{\Delta}^1,\dots, \widetilde{\Delta}^\V)$.
\end{Corollary}

{\bf Example 1}. $\Sigma = S^2$, $\Gamma$ is the graph with one edge, one vertex and
with two faces (two 1-gones which are two basic polygons) and two capitals, $c_1$, $c_2$,
(one loop with one vertex drawn on Riemann sphere). There are 
two matrices $A$ and $\hat{A}$ assigned to the vertex. And the monodromies related to the basic polygons and 
to the vertex are respectively equal to
\[
 M(c_1)=ZA,\quad M(c_2)=\hat{Z}\hat{A}\quad {\rm and}\quad {\cal{A}}=A\hat{A}
\]
where $Z$ and $\hat{Z}$ are Hermitian conjugated and where $A$ and $\hat{A}$ are two independent source matrices.

(a) The simplest example is $\Delta^1=\Delta^2=(1)$. It is the one-sheet cover $d=1$:
$$
I(\Gamma,\{A\},\Delta^1,\Delta^2)=
\int\limits_{\mathcal{Z}}\texttt{tr}\left(ZA\right)
\texttt{tr}\left(\hat{Z}\hat{A}\right)  d\Omega.
$$
 One gets
$$
I(\Gamma,\{A\},(1),(1))=
N^{-1}\sum_{a,b=1}^N A_{ab}{\hat{A}}_{ba} = N^{-1}\texttt{tr} \left(A \hat{A}  \right)
$$
As one expects,  the Hurwitz number of the one-sheeted covering is equal to 1.
 
(b) Consider the 3-sheeted covering where the profiles are chosen  to be
$\Delta^1=\Delta^2=(3)$. Then 
$$
M_{c_1}^{\Delta^1}=\frac{1}{3} \texttt{tr}\left(ZAZAZA\right),\quad
M_{c_2}^{\Delta^2}
=\frac{1}{3} \texttt{tr}\left(\hat{Z}\hat{A}\hat{Z}\hat{A}\hat{Z}\hat{A}\right)
$$
and
\[
M(\Gamma,\{ZA\},\{\Delta\},)= \frac{1}{9} \texttt{tr}\left(ZAZAZA\right)
\texttt{tr}\left(\hat{Z}\hat{A}\hat{Z}\hat{A}\hat{Z}\hat{A}\right)
\]
We have $d!=6$ systems of gluing the pairs of $Z,\hat{Z}$ in the integral
\[
I(\Gamma,\{A\},(3),(3))=\frac 19
\int\limits_{\mathcal{Z}^{3}}\texttt{tr}\left(ZAZAZA\right)
\texttt{tr}\left(\hat{Z}\hat{A}\hat{Z}\hat{A}\hat{Z}\hat{A}\right)  d\Omega =
\]
\[
 \frac 19  \int Z_{a_1b_1}A_{b_1a_2}Z_{a_2b_2}A_{b_2a_3}Z_{a_3b_3}A_{b_3a_1}  
 \hat{Z}_{\hat{a}_1\hat{b}_1}(\hat{A}_1)_{\hat{b}_1\hat{a}_2}\hat{Z}_{\hat{a}_2\hat{b}_2}
 \hat{A}_{\hat{b}_2\hat{a}_3}\hat{Z}_{\hat{a}_3\hat{b}_3}\hat{A}_{\hat{b}_3\hat{a}_1} 
 d\Omega
\]
where the summation over repeating indexes is implied.

The gluing of sheets is the same as the selection the non-vanishing types of monomials
which means the pairing the entries of $Z$ and $\hat{Z}$.
The pairings (i) of the first $Z$ from the left to the first $\hat{Z}$ from the left in the integrand,
then (ii) the second $Z$ from the left to the second $\hat{Z}$ and at last, (iii)
the third $Z$ from the left to the third $\hat{Z}$ is related, respectively, to the equalities
\[
 a_1=\hat{b}_1,\,b_1=\hat{a}_1,\qquad a_2=\hat{b}_2,\,b_2=\hat{a}_2,\qquad a_3=\hat{b}_3,\,b_3=\hat{a}_3
\]
thus, one can get rid of the subscripts with the hats.

One can record the way of gluing as $1\to 1,\,2\to 2,\, 3\to 3$.
As one can see the integral of such monomials results in $\texttt{tr}\left(A \hat{A} \right)^3$. 

There are two other ways of gluing where one obtains the same answer. One is the case where 
we glue the first $Z$ to the third $\hat{Z}$, the second to the first and the third to the second,
let us present the gluing of the sheets as $1\to 3,\,2\to 1,\, 3\to 2$, that is we put
\[
 a_1=\hat{b}_3,\,b_1=\hat{a}_3,\qquad a_2=\hat{b}_1,\,b_2=\hat{a}_1,\qquad a_3=\hat{b}_2,\,b_3=\hat{a}_2
\]
Next, we glue the first $Z$ to the second $\hat{Z}$, the second to the third and the third to the first,
let us present it as $1\to 2,\,2\to 3,\, 3\to 1$, that is we put
\[
 a_1=\hat{b}_2,\,b_1=\hat{a}_2,\qquad a_2=\hat{b}_3,\,b_2=\hat{a}_3,\qquad a_3=\hat{b}_1,\,b_3=\hat{a}_1
\]
Thus, these three ways of gluing result in $\tilde{\Delta}=(3)$.

One can see that the both ways of gluing $1\to 1,\,2\to 3,\, 3\to 2$ and 
$1\to 2,\,2\to 1,\, 3\to 3$ result in the same answer which is 
$\left(\texttt{tr}\left(A\hat{A}\right)  \right)^3$,
which is related to $\tilde{\Delta}=(1,1,1)$.

The last way of gluing $1\to 3,\,2\to 2,\, 3\to 1$:
\[
 a_1=\hat{b}_3,\,b_1=\hat{a}_3,\qquad a_2=\hat{b}_2,\,b_2=\hat{a}_2,\qquad a_3=\hat{b}_1,\,b_3=\hat{a}_1
\]
results in 
$\left(\texttt{tr}\left(A\hat{A}\right)  \right)^3 $, which is related to
the case $\tilde{\Delta}=(3)$.

In the end, we obtain 
\[
 H_2\left((3),(3),(3) \right)=\frac{3}{9}=\frac 13,\quad
 H_2\left((3),(3),(1,1,1) \right)=\frac 13,\quad
 H_2\left((3),(3),(2,1) \right)=0
\]

{\bf Example 2} $\Sigma$ is the torus. The graph $\Gamma$ consists of a single
vertex, two edges and has a single face (the single basic polygon is 4-gon).

The monodromies related to the 4-gon and to the single vertex are respectively equal to
\[
 M(c)=Z_1A_1Z_2A_2\hat{Z}_1\hat{A}_1\hat{Z}_2\hat{A}_2 \quad {\rm and}\quad{\cal{A}}=\hat{A}_2 \hat{A}_1 A_2A_1
\]
where  $Z_i,\hat{Z}_i$ is the pair of Hermitian conjugated matrices, $i=1,2$, and where $A_i,\hat{A}_i,\,i=1,2$
are four source matrices.

(a) The simplest example is $\Delta=(1)$. It is the one-sheet cover $d=1$:
$$
I(\Gamma,\{A\},(1))=
\int\limits_{\mathcal{Z}}\texttt{tr}\left(   Z_1A_1Z_2A_2\hat{Z}_1\hat{A}_1\hat{Z}_2\hat{A}_2
\right)  d\Omega
= N^{-2}\texttt{tr}\left(\hat{A}_2 \hat{A}_1 A_2A_1\right)
$$

(b) Now we choose $\Delta=(1,1)$.

$$
M^{\Delta^1}_c=\frac{1}{2} 
\left(\texttt{tr}\left( Z_1A_1Z_2A_2\hat{Z}_1\hat{A}_1\hat{Z}_2\hat{A}_2 \right)\right)^2
$$
$$
I(\Gamma,\{A\},(1,1))=\frac 12 
\int\limits_{\mathcal{Z}^{2}} 
\left(\texttt{tr}\left( Z_1A_1Z_2A_2\hat{Z}_1\hat{A}_1\hat{Z}_2\hat{A}_2 \right)\right)^2
d\Omega =
$$
$$
\frac 12 
\int\limits_{\mathcal{Z}^{2}} 
( Z_1 )_{a^{1}_1 b^{1}_1  }( A_1 )_{b^{1}_1 a^{2}_1  }
( Z_2 )_{a^{2}_1 b^{2}_1  }( A_2 )_{b^{2}_1 \hat{a}^{1}_1  }
( \hat{Z}_1 )_{\hat{a}^{1}_1 \hat{b}^{1}_1  }( \hat{A}_1 )_{\hat{b}^{1}_1 \hat{a}^{2}_1  }
( \hat{Z}_2 )_{\hat{a}^{2}_1 \hat{b}^{2}_1  }( \hat{A}_2 )_{\hat{b}^{2}_1 a^{1}_1  }
$$
$$
\times
( Z_1 )_{a^{1}_2 b^{1}_2  }( A_1 )_{b^{1}_2 a^{2}_2  }
( Z_2 )_{a^{2}_2 b^{2}_2  }( A_2 )_{b^{2}_2 \hat{a}^{1}_2  }
( \hat{Z}_1 )_{\hat{a}^{1}_2 \hat{b}^{1}_2  }( \hat{A}_1 )_{\hat{b}^{1}_2 \hat{a}^{2}_2  }
( \hat{Z}_2 )_{\hat{a}^{2}_2 \hat{b}^{2}_2  }( \hat{A}_2 )_{\hat{b}^{2}_2 a^{1}_2  }
d\Omega
$$
where the summation over repeating subscripts ${a}^{i}_j,\,{b}^{i}_j,\, \hat{a}^{i}_j,\,\hat{b}^{i}_j $ is implied,
$i=1,2$ and $j=1,2$. 

One gets
$$
\frac 12 N^{-4} 
( A_1 )_{b^{1}_1 a^{2}_1  }
( A_2 )_{b^{2}_1 \hat{a}^{1}_1  }
( \hat{A}_1 )_{\hat{b}^{1}_1 \hat{a}^{2}_1  }
( \hat{A}_2 )_{\hat{b}^{2}_1 a^{1}_1  }
( A_1 )_{b^{1}_2 a^{2}_2  }
( A_2 )_{b^{2}_2 \hat{a}^{1}_2  }
( \hat{A}_1 )_{\hat{b}^{1}_2 \hat{a}^{2}_2  }
( \hat{A}_2 )_{\hat{b}^{2}_2 a^{1}_2  }
$$
where one should equate all indexes marked with hats to indexes without hats, as well as summing over
duplicate indexes.

We have 4 ways of the equating of indexes (or, which is the same, of gluing the sheets) which are equations of type
$
 a^{i}_j=\hat{b}^i_k $ and $ b^{i}_j=\hat{a}^i_k  $ where only the subscripts $j$ and $k$ can be different.
The first one is
$$
a^{1}_1=\hat{b}^1_1 ,\, b^{1}_1 =\hat{a}^1_1 ,\quad
a^{2}_1=\hat{b}^2_1 ,\, b^{2}_1 =\hat{a}^2_1 ,\quad
a^{1}_2=\hat{b}^1_2 ,\, b^{1}_2 =\hat{a}^1_2 ,\quad
a^{2}_2=\hat{b}^2_2 ,\, b^{2}_2 =\hat{a}^2_2 
$$
thus, one can get rid of subscripts with hats.
The summation of monomials over repeating subscripts of the matrix entries yields the term
$\frac12 N^{-2}
\left(\texttt{tr}\left( \hat{A}_2 \hat{A}_1 A_2A_1\right)\right)^2$, thus it is related to
$\tilde{\Delta}=(1,1)$.
Three different ways of gluing:
$$
a^{1}_1=\hat{b}^1_2 ,\, b^{1}_1 =\hat{a}^1_2 ,\quad
a^{2}_1=\hat{b}^2_1 ,\, b^{2}_1 =\hat{a}^2_1 ,\quad
a^{1}_2=\hat{b}^1_1 ,\, b^{1}_2 =\hat{a}^1_1 ,\quad
a^{2}_2=\hat{b}^2_2 ,\, b^{2}_2 =\hat{a}^2_2 
$$
$$
a^{1}_1=\hat{b}^1_1 ,\, b^{1}_1 =\hat{a}^1_1 ,\quad
a^{2}_1=\hat{b}^2_2 ,\, b^{2}_1 =\hat{a}^2_2 ,\quad
a^{1}_2=\hat{b}^1_2 ,\, b^{1}_2 =\hat{a}^1_2 ,\quad
a^{2}_2=\hat{b}^2_1 ,\, b^{2}_2 =\hat{a}^2_1 
$$
and
$$
a^{1}_1=\hat{b}^1_2 ,\, b^{1}_1 =\hat{a}^1_2,\quad
a^{2}_1=\hat{b}^2_2 ,\, b^{2}_1 =\hat{a}^2_2 ,\quad
a^{1}_2=\hat{b}^1_1 ,\, b^{1}_2 =\hat{a}^1_1 ,\quad
a^{2}_2=\hat{b}^2_1 ,\, b^{2}_2 =\hat{a}^2_1 
$$
yields the same result. It means that
\[
 H_0\left((1,1),(2)  \right)=0,\quad  H_0\left((1,1),(1,1)  \right)=\frac 42 =2
\]

The purely geometric consideration of this answer is as follows. Find the weighted number of 
2-sheet coverings $ \varphi $  (the Hurwitz number) of the torus without branching. 
Fix a basis $ {A, B} $ of the fundamental group of the 
torus $ T $. The number of the connected components (circuits) of the inverse images $ \varphi^{- 1} (A) $ and $
\varphi^{- 1} (B) $ is 1 or 2. This gives 4 non-isomorphic covers
(one unconnected and 3 connected). The automorphism group of each of the covers is $ Z_2 $. 
Thus, the Hurwitz number is 2.

(c) Now we choose $\Delta=(2)$.

$$
M^{\Delta^1}_c=\frac{1}{2} 
\texttt{tr}\left(\left( Z_1A_1Z_2A_2\hat{Z}_1\hat{A}_1\hat{Z}_2\hat{A}_2 \right)^2\right)
$$
$$
I(\Gamma,\{A\},(2))=\frac 12 
\int\limits_{\mathcal{Z}^{2}} 
\texttt{tr}\left(\left( Z_1A_1Z_2A_2\hat{Z}_1\hat{A}_1\hat{Z}_2\hat{A}_2 \right)^2\right)
d\Omega =
$$
$$
\frac 12 
\int\limits_{\mathcal{Z}^{2}} 
( Z_1 )_{a^{1}_1 b^{1}_1  }( A_1 )_{b^{1}_1 b^{1}_1  }
( Z_2 )_{a^{2}_1 b^{2}_1  }( A_2 )_{b^{2}_1 b^{2}_1  }
( \hat{Z}_1 )_{\hat{a}^{1}_1 \hat{b}^{1}_1  }( \hat{A}_1 )_{\hat{b}^{1}_1 \hat{b}^{1}_1  }
( \hat{Z}_2 )_{\hat{a}^{2}_1 \hat{b}^{2}_1  }( \hat{A}_2 )_{\hat{b}^{2}_1  a^{1}_2 }
$$
$$
\times( Z_1 )_{a^{1}_2 b^{1}_2  }( A_1 )_{b^{1}_2 b^{1}_2  }
( Z_2 )_{a^{2}_2 b^{2}_2  }( A_2 )_{b^{2}_2 b^{2}_2  }
( \hat{Z}_1 )_{\hat{a}^{1}_2 \hat{b}^{1}_2  }( \hat{A}_1 )_{\hat{b}^{1}_2 \hat{b}^{1}_2 }
( \hat{Z}_2 )_{\hat{a}^{2}_2 \hat{b}^{2}_2  }( \hat{A}_2 )_{\hat{b}^{2}_2 a^{1}_1  }
d\Omega
$$
One gets the 4 sets of equations identical to the previous case, however, now the result is
$\frac12 N^{-4}
\texttt{tr}\left( \left(\hat{A}_2 \hat{A}_1 A_2A_1\right)^2\right)$ which is related to
$\tilde{\Delta}=(2)$.
\[
 H_0\left((2),(1,1)  \right)=0,\quad  H_0\left((2),(2)  \right)=\frac 42 =2
\]

\subsection{Non-orientable case. Integrals and the topological theory \label{non-orientable}}

Hurwitz numbers for non-orientable surfaces were studied in 
\cite{M2},\cite{GARETH.A.JONES},\cite{AN1},\cite{AN2008}. In the context of integrable
systems it was done in \cite{NO-2014},\cite{NO-LMP},\cite{Carrell}.
Now we relate it to matrix integrals.

Consider
\be\label{Mob-tau}
\tau^{\rm Mobius}\left(M(c_i)\right) =\sum_\lambda s_\lambda(M(c_i)) = 
\sum_{\Delta} D(\Delta) M^{\Delta}_{c_i}
\ee
where $D(\Delta)$is given by (\ref{D(Delta)}),
and
\be\label{handle-tau}
 \tau^{\rm handle}\left(M(c_i),M(c_j)\right)=
 \sum_\lambda s_\lambda(M(c_i))s_\lambda(M(c_j))
 =\sum_{\Delta} 
 z_\Delta M^{\Delta}_{c_i} M^{\Delta}_{c_j}
\ee
where sums range over all partitions. 
We use  (\ref{Mob-tau}) for the description of (\ref{Moebius-cut}) and we use
(\ref{handle-tau}) in relation with (\ref{Klein-cut}) and (\ref{handle}), see Figure \ref{f3},
as follows:

\begin{Theorem}\label{integral=handle-cuts}
\[
 \int 
 \left(\prod_{i=1}^{\textsc{h}}
 \tau^{\rm handle}\left(M(c_{2i}),M(c_{2i-1}\right)\right)
 \left(\prod_{i=2\textsc{h}+1}^{2\textsc{h}+\textsc{m}}\tau^{\rm Mobius} (M(c_i))\right)
 \left(\prod_{i=2\textsc{h}+\textsc{m}+1}^{\f}M^{\Delta^i}_{c_i}\right)
 d\Omega 
 \]
 \[
 =N^{-nd}\sum_{\tilde{\Delta}^1,\dots,\tilde{\Delta}^\V}
 {\cal{A}}_{w_1}^{\widetilde{\Delta}^1}
\dots {\cal{A}}_{w_V}^{\widetilde{\Delta}^\V}
 H_{\tilde{\Sigma}}\left(\Delta^{\f-\textsc{m}-2\textsc{h}+1},\dots,\Delta^{\f},
 \tilde{\Delta}^1,\dots,\tilde{\Delta}^\V
 \right)
\]
where the Euler characteristic of $\tilde{\Sigma}$ is equal to $\f-n+\V-\textsc{m}-2\textsc{h}$.
\end{Theorem}
The interpretation of the equality in the Theorem can be as follows. Consider the original orientable surface
$\Sigma$ and the drawn graph $\Gamma$ of Subsection \ref{Geom comb Hurwitz}  and remove $\textsc{h}$ pairs of faces with capitals 
$\left(c_{2i},c_{2i-1}\right),\,i=1,\dots,2\textsc{h}$ and seal each pair of  holes with a handle. 
Further, in addition,  remove
$\textsc{m}$ faces with capitals $c_i,\,i=2\textsc{h}+1,\dots,2\textsc{h}+\textsc{m}$ and seal each one 
with a Moebius strip. We obtain $\tilde{\Sigma}$.
\begin{proof}
Proof follows from the axioms of topological theories
presented by relations (\ref{handle-cut}) and (\ref{Hurwitz-down}) and
from the explicit form of (\ref{Mob-tau}) and of (\ref{handle-tau}) and  from Theorem \ref{Theorem}.
\end{proof}

\section{Hurwitz numbers and quantum and classical integrable models \label{Hurwitz-quantum-models}}

In this section we show the links of Hurwitz numbers with certain quantum integrable systems.

\subsection{Dubrovin's commuting quantum Hamiltonians, Okounkov's
completed cycles, Jucys-Murphy elements, classical integrable systems \label{Dubr-quant}}

\paragraph{Vertex operators of Kyoto school.} Consider one-parametric series of 
differential operators in infinitely many variables
$\bpow =(p_1,p_2,\dots)$:
\be\label{FourierBoson}
\theta(z)=\sum_{m>0} \frac 1m z^m p_m  + p_0 \log z  - 
\sum_{m>0} z^{-m} \frac{\partial}{\partial p_m},\quad |z|=1
\ee
which depends on the parameter $z\in S^1$ and acts on formal series in $p_1,p_2,\dots$.

Properly ordered exponentials of $\theta(z)$ and properly defined products of such exponentials  
are known in physics as vertex  operators\footnote{ 
At first, the vertex operators were considered in the work of Pogrebkov and Sushko
in \cite{PogrebkovSushko} in their studies of quantum Thirring model. }.
As it was explained in the works of Kyoto school, the vertex operators
play an important role in the theory of classical integrability.

Now, let us consider
the vertex operator that
depends on two parameters $z$ and ${\q}=e^y$:
\[
 ::e^{\theta(z{\q}^{1/2})-\theta(z{\q}^{- 1/2})}  ::=\q^{ p_0}
 e^{\sum_{m>0}\frac 1m  z^m\left( \q^{\frac m2} - \q^{-\frac m2}  \right)p_m } 
e^{\sum_{m>0} z^{-m} \left( \q^{\frac m2} - \q^{-\frac m2}  \right)   
\frac{\partial}{\partial p_m}}
\]

Dots $::A::$ denote the so-called bosonic normal ordering of $A$, that means that all differential operators 
$\frac{\partial}{\partial p_m},\,m>0$ called in physics "ellimination operators" which are present in $A$ are moved 
to the right, while all "creation operators", which
are all $p_m,\,m\ge 0$, are moved to the left.

This vertex operator acts on functions of $p_1,p_2,\dots$ as a shift operator in infinitely many variables.
In classical soliton theory such vertex operator acts on the so-called tau functions \cite{JM} where this 
action means the adding of the so-called soliton to a solution to Kadomtsev-Petviashvili equation 
\cite{NovikovManakovZakharov}. In \cite{Orl1987}, \cite{GrinevichOrlov} it was used to construct
symmetries of the KP hierarchy sometimes called $\hat{W}_{1+\infty}$ symmetries.

\paragraph{Quantum Hamiltonians.} Hamiltonians of simple quantum models can be obtained as follows.
Consider one parametric family of vertex operators
\be\label{Ham(q)}
{\hat H(\q)}:= \res_z 
\frac { ::e^{\theta(z{\q}^{1/2})-\theta(z{\q}^{- 1/2})}  :: -1 }{{\q}^{ 1/2}-{\q}^{-1/2}} \frac{dz}{z}
=  \sum_{n>0} y^n {\hat H}_n
\ee
where $\q=e^y \in \mathbb{C}$.

The important fact pointed out in \cite{Okounkov-Pand-2006} (see also \cite{MM2},\cite{MM3}) is
that $H_3$ is the cut-and-join operator (\ref{Hopf}), introduced in \cite{GJ}.

\bp For any $\q_1$ and $\q_2$
\[
 [ {\hat H(\q_1)}, {\hat H(\q_2)}   ] =0,\quad [ {\hat H}_n, {\hat H(\q_2)}   ] =0
\]
\[
[ {\hat H_n}, {\hat H_m}   ] =0
\]
\ep
This fact is well known in soliton theory, see for instance \cite{Orl1987}.

Next, let us recall the definition of Schur function. Consider the equality
\[
 e^{\sum_{m>0} \frac 1m z^m  p_m } = 1+z p_1+z^2\frac{p_1^2+p_2}{2}+\dots =\sum_{m\ge 0} z^m s_{(m)}(\bpow)
\]
The polynomials $s_{(m)}$ are called elementary Schur functions. Then, the polynomials
$s_\lambda$ labeled by a partition $\lambda=(\lambda_1,\lambda_2,\dots)$ defined by
\[
 s_\lambda(\bpow) = \det\left( s_{(\lambda_i-i+j)}(\bpow)  \right)_{i,j\ge 1}
\]
are called Schur functions. (In the right hand side, it is implied that $s_{(m)}=0$ for negative $m$.)
The Schur functions form a basis in the space of polynomial functions (for details see the textbook \cite{Mac}).

A quantum integrable model can be defined as a set of commuting operators (Hamiltonians),
the linear space where these operators act (the Fock space), and a special vector of the Fock space which
is eliminated by the Hamiltonians (the so-called vacuum vector).
The dynamical Hamiltonian equation is obtained by the replacing of the Poisson bracket by the commutator
and is called the Heisenberg equation in physics.

Let us introduce
\[
 \theta(z,\bt)=e^{\sum_{m>0} \frac 1m t_m\hat{H}_m}\theta(z)e^{-\sum_{m>0} \frac 1m t_m\hat{H}_m}
\]
where the set $\bt=(t_1,t_2,\dots)$ is the set of parameters called higher times which are
the analogue of the higher times in classical integrable models.

\bp \label{Dubrovin} (Dubrovin \cite{Dubr}). 

(a) The set of commuting operators $\{ H_n \}$ can be interpreted as the 
set of commuting Hamiltonians of the set of quantum dispersionless KdV equations 
(or, that is the same, the hierarchy of quantum dispersionless equations) on the circle $|z|=1$:
\be\label{quantum-dKdV}
 \frac{\partial {\hat u}}{\partial t_n} = [{\hat H}_n,{\hat u}],\quad n=1,2,\dots
\ee
where ${\hat u}(z,\bt)=\sqrt{-1}z\frac{\partial\theta(z,\bt)}{\partial z}$. The Fock space
of the set of quantum dispersionless KdV equations is the space of polynomial functions
in the variables $p_1,p_2,\dots$.

(b) The Hamiltonian ${\hat H_2}$ which is the Hamiltonian of the quantum dispersionless
KdV equation itself coincides with the cut-and-join operator (\ref{Hopf}).

(c) For any $\lambda$, Schur functions $s_\lambda(\bpow)$ are eigenfunctions of these Hamiltonians, or, the same,
the Schur functions are eigenstates of quantum dispersionless KdV equations:
\[
 {\hat H}_n s_\lambda(\bpow) = \left(\left(\frac 12 + \lambda_i-i  \right)^n 
  - \left(\frac 12 -i  \right)^n\right) s_\lambda(\bpow)
\]
The vacuum vector is given by $s_0=1$.

\ep

The proof of this statement is contained in \cite{Dubr} (see also \cite{E}). (The last formula was known much earlier
from the boson-fermion correspondence\footnote{
As it was noted in \cite{Orl1987},
the  existence of commuting operators obtained from
the expansion of vertex operators gives rise to the commuting symmetries and additional hierarchies
of commuting flows compatible with the KP flows. The classical Burgers equation
is the example, this fact was noted in \cite{Zakharov}. Then, Dubrovins dispersionless KdV is the quantization
of this classical dispersionless KdV.}.)

\br 
 Hamiltonians $\{{\hat H}_m \}$ belong to the Cartan subalgebra of
$\hat{W}_{1+\infty}$ and the Schur functions are eigenfunctions of the whole set
of these operators. There exits
 more general algebra of commuting operators 
 which does not belong to $\hat{W}_{1+\infty}$ and whose eigenfunctions  are the Schur functions,
 see  \cite{MM1},\cite{MM3}.
\er

Now, let us treat the generating function for the KdV Hamiltonians (\ref{Ham(q)}) as the Hamiltonian that depends on
the parameter $\q$. The related quantum equation is a $\q$-version of the quantum dispersionless 
Toda lattice equation\footnote{We note that this quantum model is related to the so-called free fermion point.}:
\be\label{toda-free}
 \frac{\partial^2\theta(z,{\bf t})}{\partial z \partial t} 
=::e^{\theta(z)-\theta(z{\q}^{- 1})}::- ::e^{\theta(z{\q})-\theta(z)}::
 \ee
(The right hand side is obtained as $[\theta_z(z,{\bf t}),{\hat H({\q})}] $). Due to
$[{\hat H}_m,H(\q)]=0$ this evolutionary equation  is compatible with the quantum
dispersionless KdV equations (\ref{quantum-dKdV}), that means 
$\partial_t\left(\partial_{t_m}\theta_z\right)=\partial_{t_m}\left(\partial_{t}\theta_z\right)$.
 
 The expansion of the right hand side of (\ref{toda-free}) in $y=\log\q$ gives rise to the higher
dispersionless quantum KdV equations.

 From Proposition \ref{Dubrovin} it follows
\begin{Corollary} 
\label{Lemma-Okounkov-Dubrovin}
\cite{Okounkov-Pand-2006} For any Young diagram $\lambda$, the Schur function $s_\lambda$ is 
the eigenfunction of the operator $\hat{H}(\q)$
\be
\hat{H}(\q) s_\lambda(\bpow) =  {\bf e}_\lambda({\q})s_\lambda(\bpow)
\ee
\be\label{E}
{\bf e}_\lambda({\q})=\sum_{i=1}^{\infty} \left({\q}^{\frac 12 +\lambda_i-i }-\q^{\frac 12 -i}\right)
\ee
\end{Corollary}
Formula (\ref{E}) was used by Okounkov to generate the so-called completed cycles,
see \cite{Okounkov-Pand-2006}.

\bl\label{JMcontent-q}
\be
 e^{t(\q^{\frac 12}-\q^{-\frac 12})\sum_{i=1}^d \q^{\gJ_i}} \gF_\lambda = 
 e^{t{\bf e}_\lambda(\q)} \gF_\lambda
\ee
\el

Proof follows from (\ref{Jucys-Murphy-content}) and from the relation
between the eigenvalues ${\bf e}_\lambda({\q})$ and the so-called quantum contents $\q^{i-j}$
of the nodes of Young diagram $\lambda$ with coordinates $i,j$:
 \be\label{T-m}
\sum_{(i.j)\in\lambda} \q^{j-i} =
 \frac{ {\bf e}_\lambda({\q}) }{\q^{\frac 12}-\q^{-\frac 12}}
 \ee
 The last equality is obtained by the re-summation of the left hand side.

Let us introduce the following Hurwitz numbers
\be
H_\Sigma(\Delta^1,\dots,\Delta^k;t)=<e^{t(\q^{\frac 12}-\q^{-\frac 12})
\sum_{i=1}^d \q^{\gJ_i}}\gC_{\Delta^1}\cdots \gC_{\Delta^k}>_\Sigma
\ee
where $\gJ_i$ are Jusys-Murphy elements, see Subsection \ref{Hurwitz numbers and representation theory}.

Let us mention that the expansion of the left side in the small parameter $y=\log\q$ generates Hurwitz 
numbers defined on the completed cycles.

\bp We have
\be\label{t-Hurwitz-q}
e^{t\hat{H}(\q)} \int e^{\sum_{m>0}\frac 1m p_m M(c_1)}\prod_{i>1}^\f M^{\Delta^i}(c_i) d\Omega
=\sum_{\tilde{\Delta}^1,\dots,\tilde{\Delta}^\V}  
H_\Sigma(\Delta^1,\dots,\Delta^\f,\tilde{\Delta}^1,\dots,\tilde{\Delta}^\V;t)\bpow_{\Delta^1}
\prod_{i=1}^\V  {\cal{A}}^{\tilde{\Delta}^i}_i
\ee
\ep

Proof follows from Cauchy-Littlewood identity
\[
 e^{\sum_{m>0}\frac 1m p_m M(c_1)}=\sum_\lambda s_\lambda(\bpow)s_\lambda(M(c_1)),
\]
from Corollary \ref{Lemma-Okounkov-Dubrovin} and from Lemma \ref{JMcontent-q}.

\subsection{Hurwitz numbers (\ref{t-Hurwitz-q}) generated by tau functions. \label{Dubrovin2}}

Let us consider the commuting set of quantum Hamiltonians 
$\hat{H}(\q),\hat{H}(\q^2),\hat{H}(\q^3),\dots$ which generate the following evolution flows
\[
 \theta(z,\bt(\q))=e^{\sum_{m>0} \frac 1m t_m(\q)\hat{H}(\q^m)}\theta(z)e^{-\sum_{m>0} \frac 1m t_m(\q)\hat{H}(\q^m)}
\]

The flow parameters $\bt(\q)=(t_1(\q),t_2(\q),\dots ) $ can be treated as the higher times for \textit{soliton 
lattice} solutions related to the Hurwitz problem. In this subsection we shall use the notation $t_m$ instead
of the notation $t_m(\q)$ and the notation $\bt$ instead of $\bt(\q)$.

We refer the reader to
the important notion of \textit{tau function}  introduced by Kyoto school, see \cite{JM} for a review of the topic.
There are only few cases where Hurwitz numbers (\ref{t-Hurwitz-q}) are generated by tau functions and we are going
to select them. Let us note that the tau functions we need is are the tau functions of the hypergeometric family
described in \cite{O-TMP-2006} (the so-called 2KP case) and in \cite{OST-I} (the so-called BKP case).
As functions of the variables
$\bpow$ 2KP tau functions are written as series in the Schur functions 
( \cite{Takasaki-Schur},\cite{TI}
and also in \cite{KMMM},\cite{OS-TMP} and the specification we need in is written down in \cite{O-TMP-2006}. 
These tau functions
solve a pair KP hierarchy: one KP hierarchy with respect to the higher times $\bpow=(p_1,p_2,\dots )$ and other KP
hierarchy with respect to the higher times $\bt=(t_1,t_2,\dots)$. Below we also need the BKP analogue of KP series written down in \cite{OST-I}. 
As functions in the variables $\bt$
solutions below presented in \cite{O-TMP-2006} and in \cite{OST-I} may be called soliton lattices 
in the $\mathbb{R}^\infty$ space with coordinates $\bt$.

\bl  Let $\bt=(t_1,t_2,\dots)$, $k\in\mathbb{Z}$ is the infinite set of parameters and 
$\bpow(0,\q)=\left(p_1(0,\q),p_2(0,\q),\dots \right)$ is the  special 
set given by $p_m(0,\q)=\frac{1}{1-\q^m}$.
Then
\be
\tau_k(\bpow,\bt)=\sum_{\lambda} e^{\sum_{m>0}\frac{t_m\q^{km}}{m}
\left(\q^{m(\frac 12 +\lambda_i-i)} -\q^{m(\frac 12 -i)}\right)}
s_\lambda(\bpow)s_\lambda(\bpow(0,\q))
\ee
\[
=e^{\sum_{m>0} \frac 1m t_m\q^{mk}\hat{H}(\q^m)}\cdot 
e^{\sum_{m>0}\frac 1m \frac{p_m}{1-\q^m}}
\]
is a KP tau function (the soliton lattice one) with respect to the parameters $\bt$ which plays the role of the KP
higher times and the parameters $\bpow$ and $k$ define initial configuration of solitons.
It is also a KP tau function of hypergeometric family with respect to the parameters $\bpow$ which, now, 
play the role of the KP higher times
while the parameters $\bt$ and $k$ are given constants.
It is also a tau function of the two-dimensional Toda lattice hierarchy with respect to the sets $\bt$, $\bpow$ and 
the discrete variable $k$. The last statement means that 
$\phi_k(\bpow,\bt)=\log\tau_k(\bpow,\bt)-\log\tau_{k-1}(\bpow,\bt)$ solves Toda lattice equation
$\partial_{t_1}\partial_{p_1}\phi_k=e^{\phi_{k-1}-\phi_{k}}-e^{\phi_{k}-\phi_{k+1}}$.
\el

The proof is basically contained in \cite{O-TMP-2006}.
\bp\label{Prop-Jucys-Murphy-tau-TL}
\be
\tau_k(\bpow,\bt)=
\sum_{\Delta^1,\Delta^2} H
_{S^2}\left(\Delta,\Delta^1;\bt,k,\q\right)\bpow_{\Delta^1}\prod_{i>0} \frac{1}{1-\q^{\Delta_i^2}}
\ee
where sum ranges over all partitions $\Delta^i=(\Delta^i_1,\Delta^i_2,\dots ),\,i=1,2$ where
it is implied that $|\Delta^1|=|\Delta^1|$ and where
\be
H_{S^2}(\Delta^1,\Delta^2;\bt,k,\q)=
<e^{\sum_{m>0}t_m\q^{km}\frac {\q^{\frac m2}-\q^{-\frac m2}}{m} 
\sum_{i=1}^d \q^{m\gJ_i}}\gC_{\Delta^1} \gC_{\Delta^2}>_{S^2}
\ee
\ep

Similarly, we have

\bl  Let $\bt=(t_1,t_2,\dots)$, $k\in\mathbb{Z}$ is the infinite set of parameters and
$\bpow(0,\q)=\left(p_1(0,\q),p_2(0,\q),\dots \right)$ is the  special 
set given by $p_m(0,\q)=\frac{1}{1-\q^m},\,m>0$.
Then
\be
\tau_k^{\rm B}(\bt)=\sum_{\lambda} e^{\sum_{m>0}\frac{t_m\q^{km}}{m}
\left(\q^{m(\frac 12 +\lambda_i-i)} -\q^{m(\frac 12 -i)}\right)}
s_\lambda(\bpow(0,\q))
\ee
\[
=\left[ e^{\sum_{m>0} \frac 1m t_m\q^{mk}\hat{H}(\q^m)}\cdot 
\sum_{\lambda} s_\lambda(\bpow)\right]_{\bpow=\bpow(0,\q)}
\]
is an infinite-soliton BKP tau function presented in \cite{OST-I} with respect to the parameters $\bt$ and $k$ 
which play the role of the BKP higher times. 
\el

The proof is basically contained in \cite{OST-I}.
\bp\label{Prop-Jucys-Murphy-tau-BKP}
\be
\tau_k^{\rm B}(\bt)=
\sum_{\Delta} H_{\mathbb{RP}^2}\left(\Delta;\bt,k,\q\right)
\prod_{i>0} \frac{1}{1-\q^{\Delta_i}}
\ee
where sum ranges over all partitions $\Delta^i=(\Delta^i_1,\Delta^i_2,\dots ),\,i=1,2$ where
\be
H_{\mathbb{RP}^2}(\Delta;\bt,k,\q)=
<e^{\sum_{m>0} {t_m\q^{km}}\frac{ \q^{\frac m2}-\q^{-\frac m2} }{m}
\sum_{i=1}^d \q^{m\gJ_i}}\gC_{\Delta}>_{\mathbb{RP}^2}
\ee
\ep

\br\label{Jucys-Murphy-tau}
Simirlaly, we can show that
$\frac{\partial}{\partial t_1}\log < e^{\sum_{i=1}^d\sum_{m>0}\frac 1m t_m^* \gJ_i^m}  >_{S^2}$
solves KP equation with respect to the variables $t_1^*,t_2^*,t_3^*$ and according to \cite{Zakharov}  it 
also solves the so-called Burges equation. Also
$ < e^{\sum_{i=1}^d\sum_{m>0}\frac 1m t_m^* \gJ_i^m}  >_{\mathbb{RP}^2}$
is the tau function of the BKP hierarchy. We will not explain it in  details, let us point out
that in this case, one needs an additional triangle transform of from the set of the KP (respectively, of the BKP)
higher times to the set $t_1^*,t_2^*,\dots$. This can be derived from
 the relation (\ref{Jucys-Murphy-content}) and the results of \cite{NO-LMP}.

\er

\subsection{$2D$ Yang-Mills quantum theory \label{Witten-quant}}

If we slightly modify the integrand in Theorem \ref{Theorem}, then the 
integral obtained will lead to another interesting problem. This is the 
problem of calculating the correlation functions of two-dimensional quantum 
gauge theories in the case of both orientable and non-orientable surfaces.
Quantum two-dimensional gauge theories (also known as 2D Yang-Mills (2DYM) models) were analyzed in the works of 
A. Migdal  and his students \cite{Migdal},\cite{Rusakov}, and in more detail 
in Witten's work \cite{Witten}, in which the non-orientable case was also 
considered.
 The following answer for the partition function $Z_\Sigma$ of the 
two-dimensional Yang-Mills theory with the unitary gauge group was obtained
\cite{Witten}:
\[
Z_\Sigma(\rho)= 
\sum_{d\ge 0}
\sum_{\lambda\in\Upsilon_d}
e^{-\rho c_2/2}
\left({\rm dim}_G\lambda\right)^{\e}
\]
Here $\rho$ is the coupling constant of the theory and for the case $G=SU(N)$ one 
has $c_2=(\lambda_i-i+N)^2$, ${\rm dim}_G\lambda=s_\lambda(\mathbb{I}_N)$  is the dimension of the representation 
$\lambda$, $\e$ is the Euler characteristic of the surface $\Sigma$ (in \cite{Migdal} only 
the case $\Sigma=\mathbb{S}^2$ was studied).

The following answer for the correlation function of the pair of the 
so-called Wilson loops around points $z_1,z_2\in\Sigma$ was obtained
\[
 Z_{\Sigma}(\rho;\Theta_1,\Theta_2)=
\]
\be\label{Migdal}
 \left<e^{\oint_{z_1} Adz}\, e^{\oint_{z_2} Adz} \right>_{2DYM}=
 \sum_{d\ge 0}
\sum_{\lambda\in\Upsilon_d} e^{-\rho c_2/2}s_\lambda(\Theta_1)s_\lambda(\Theta_2)
\ee
The meaning of the left side of (\ref{Migdal})  is explained in \cite{Migdal}, $s_\lambda$ on the right side denotes the Schur function and
$\Theta_i\in SU(N)$ denotes the class of the Wilson loop $e^{\oint_{z_i}Adz},\,i=1,2$, $A$ is the so-called 
gauge field on $\Sigma$, which is a flat connection on $\Sigma$ without $z_1,z_2$.

In \cite{O-TMP-2006} it was noted that the right-hand side coincides with a certain hypergeometric tau 
function of the Toda lattice \cite{KMMM},\cite{OS-2000},\cite{OS-TMP}. 

The answer for the correlation function of $k$ Wilson loops of 
2DYM theory on a surface $\Sigma$ with an Euler characteristic $\e=2-2g$ 
(or, that is the same,  the Yang-Mills path integral 
over all flat connections on $\Sigma\backslash \{z_i\}$ where the holonomies around $k$ marked points 
belong to the conjugacy classes 
$\Theta_1,\dots,\Theta_k$)
is 
\cite{Witten}:
\be\label{Witten}
Z_{\Sigma}(\rho;\Theta_1,\dots,\Theta_k)=\sum_{d\ge 0}
\sum_{\lambda\in\Upsilon_d} e^{-\rho c_2/2}
\left({\rm dim}_G\lambda\right)^{\e-k}\prod_{i=1}^{k} s_\lambda(\Theta_i)
\ee

Our claim is to
 present the same right hand side of (\cite{Witten}) with the help of Gaussian integral over complex matrices
 and relate it to the counting of Hurwitz numbers.
 
\paragraph{Gaussian integrals.} 

Denote $s_\lambda(X):=s_\lambda(\bpow(X))$, where 
$\bpow(X)=(p_1(X),p_2(X),p_3(X),\dots)$ where $p_i(X)=\texttt{tr}(X^i)$.
Introduce $\bpow_\Delta:=p_{\Delta_1}p_{\Delta_2}\cdots$, where $\Delta=(\Delta_1,\Delta_2,\dots)$ 
is a given Young diagram, $|\Delta|=d$.

In what follows we need characteristic map relation (\cite{Mac}) 
\be
\label{Schur-char-map'-opposite}
\bpow_\Delta =
\frac{\operatorname{dim}\lambda}{d!}\,\sum_{\lambda\in \Upsilon_d } 
\varphi_\lambda(\Delta)s_\lambda(\bpow)
\ee

Let us recall the well-known fact that 
\be\label{dim-dim}
{\rm dim}_U\lambda = (N)_\lambda{\rm dim}\lambda,
\quad
 (N)_\lambda := \prod_{(i,j)\in\lambda} \left( N+j-i\right)
\ee

With the help of (\ref{orth1}) and (\ref{Mednykh})  one gets the following corollary of Theorem \ref{Theorem}:

\begin{Lemma}
\label{Schur-expectation}
Consider the graph $\Gamma$ the sets of monodromies $M(c_1),\dots, M(c_\f)$ and
${\cal{A}}_{w_1},\dots,{\cal{A}}_{w_V}  $
introduced
in Section \ref{Geom comb Hurwitz}.
For a given set of partions $\lambda^1,\lambda^2,\dots,\lambda^\f =\lambda$, we have
\be
 \int \left( \prod_{i=1}^\f s
_{\lambda^i}\left( M(c_i) \right) \right)d\Omega  =
\delta_{\lambda^1,\lambda^2,\dots,\lambda^\f} \left( \frac{(N)_\lambda}{N^{|\lambda|}} \right)^n
\left(s_\lambda(\mathbb{I}_N) )  \right)^{-n}  \prod_{v=1}^{\V}
s_{\lambda}\left({\cal{A}}_{w_v}\right)
\ee
  where  $ \delta_{\lambda^1,\lambda^2,\dots,\lambda^\f} $ is equal to 1 in case 
$\lambda^1=\lambda^2=\cdots =\lambda^\f$ and to 0 otherwise.
\end{Lemma}

Consider the following series over Young diagrams:
\be\label{a1}
\tau(X_1,X_2,t,a):=
\sum_{d\ge 0}
\sum_{\lambda\in\Upsilon_d}
\left(\frac{N^{|\lambda|}}{(N)_\lambda }\right)^a
e^{\frac t2 \sum_{i=1}^N(\lambda_i-i+N+\frac 12 )^2}
s_\lambda(X_1)s_\lambda(X_2)
\ee
\be\label{a2}
\tau^{\rm B}(Y,t,a) :=
\sum_{d\ge 0}
\sum_{\lambda\in\Upsilon_d}
\left(\frac{N^{|\lambda|}}{(N)_\lambda }\right)^a
e^{\frac t2 \sum_{i=1}^N(\lambda_i-i+N+\frac 12 )^2}
s_\lambda(Y)
\ee
where $a\in \mathbb{C} $, and $X_1,X_2,Y \in \mathbb{GL}_N(\mathbb{C})$.
According to \cite{OS-TMP}, (\ref{a1}) is the tau function of the Toda lattice (TL) hierarchy introduced 
in \cite{JM}.  
The sets $\texttt{tr}(X_i^m),\,m=1,2,\dots$, $i=1,2$ play the role of the higher times of the
TL hierarchy. 
According to \cite{OST-I},  (\ref{a2}) is the  tau function of the so-called BKP hierarchy introduced 
in \cite{KvdLbispec}. The set $\texttt{tr}(Y^m),\, m=1,2,\dots$ plays the role of the set of the higher 
times of the BKP hierarchy.

We get
\bp\label{int-tau=Witten} Let $a_1+\dots +a_\f = n$, $t_{1}+\cdots +t_{\f} =
{\rho}$. Suppose 
${\cal{A}}_{w_1},\dots,{\cal{A}}_{w_\V},\Theta_{2\textsc{h}+\textsc{m}+1},\dots,\Theta_{\f} \in SU(N) $
is a given set of matrices. Then
\[
 \int 
 \left(\prod_{i=1}^{\textsc{h}}
 \tau^{\rm A}\left(M(c_{2i}),M(c_{2i-1}),t_{2i}+t_{2i+1},a_{2i}+a_{2i+1}\right)\right)
 \left(\prod_{i=2\textsc{h}+1}^{2\textsc{h}+\textsc{m}}\tau^{\rm B} (M(c_i),t_i,a_i)\right)\times
 \]
 \be\label{rhs}
 \left(\prod_{i=2\textsc{h}+\textsc{m}+1}^{\f}\tau^{\rm A}\left(M(c_i),\Theta_i,t_i,a_i\right) \right)
 d\Omega 
= \sum_{\lambda} e^{-\rho c_2/2}\left(\dim_U(\lambda)\right)^{-n}\prod_{i=1}^{\f_1}s_\lambda(\Theta_i)
\prod_{v=1}^{\V}s_\lambda\left( {\cal{A}}_{w_v}\right)
\ee
where the integrand is given by formulas (\ref{a1})-(\ref{a2}). The right hand side of (\ref{rhs})
is equal to the right hand side of (\ref{Witten}) where $\Sigma$ is a surface with Euler characteristic
equal to $\f-n+\V-2\textsc{h}-\textsc{m}$, where $k=\f_1+\V$ and where the role of the classes plays the classes 
of the matrices
${\cal{A}}_{w_1},\dots,{\cal{A}}_{w_\V},\Theta_{2\textsc{h}+\textsc{m}+1},\dots,\Theta_{\f}$.
\ep

\begin{proof}
 We integrate the products of the right hand sides of (\ref{a1}) and (\ref{a2}) using
Lemma \ref{Schur-expectation} and use orthogonality relations (\ref{orth1}),(\ref{orth2}).
 
\end{proof}

\paragraph{Relation to Hurwitz numbers.}
The comparance with Theorem \ref{integral=handle-cuts} and also with the help of (\ref{Mednykh}) and character 
map relation
(\ref{Schur-char-map'-opposite}) and with the fact that
${\rm dim}\lambda$ is equal to ${\rm dim}_U\lambda$ times a polynomial in $N$ shows that right hand sides of
 (\ref{rhs}), (\ref{Witten})
generates a specific series in
Hurwitz numbers. We will do it similarly it was done in \cite{O-TMP-2017}, or  
in \cite{HarnadOverview}. 

Namely, we need
\be\label{char-map}
s_\lambda(\mathbb{I}_N)=
\frac{{\rm dim}\lambda}{d!}N^d \left(1+\sum_{d > k > 0 } \phi_\lambda(k)N^{-k}\right)\,,\quad d=|\lambda|
\ee
where
\be\label{phi}
\phi_\lambda(k)\,:=\,\sum_{\Delta\atop \ell(\Delta)=d-k} \,\varphi_\lambda(\Delta),\quad k=0,\dots,d-1
\ee
The corollary of (\ref{char-map}) is
\be
\left( s_\lambda(\mathbb{I}_N)\right)^{\e} =
  \left(\frac{{\rm dim}\lambda}{d!}\right)^{\e} N^{\e d} \left(1+\sum_{k >0}{\tilde\phi}_\lambda(k;\e ) N^{-k}   \right)
\ee
where each ${\tilde\phi}_\lambda(k)$ is built of a collection $\{ \phi_\lambda(i)\,,\,i>0 \}$:
\be\label{tilde-phi}
{\tilde\phi}_\lambda(k;\e)=\sum_{l\ge 1}\, \e(\e-1)\cdots (\e-l+1)\,
\sum_{\mu\atop \ell(\mu)=l,\,|\mu|=k}\frac{\phi_\lambda(\mu)}{|{\rm aut}\,\mu |}\,,
\quad \phi_\lambda(\mu):=\phi_\lambda(\mu_1)\cdots \phi_\lambda(\mu_l) 
\ee
Here $\mu=(\mu_1,\dots,\mu_{l'})$ is a partition which may be written alternatively \cite{Mac} as 
$\mu=\left( 1^{m_1}2^{m_2}3^{m_3} \cdots\right)$
where $m_i$ is the number of times a number $i$ occurs in the partition of $|\mu|=k$. Thus the set  of all 
non-vanishing 
$m_{j_a}\,,a=1,\dots l'$, ($l'\le l$)
defines the partition $\mu$ of length $\ell(\mu)=\sum_{a=1}^{l'} m_{j_a} = l$ and of weight 
$|\mu|=\sum_{a=1}^{l'} j_a m_{j_a} =k$. Then
the order of the automorphism group of the partition $\mu$ is
\[
  |{\rm aut}\,\mu | := m_{j_1}!\cdots m_{j_{l'}}!
\]

We see ${\tilde\phi}_\lambda(k;1)={\phi}_\lambda(k)$. Note that 
$\phi_\lambda(1)=\varphi_\lambda\left((2,1^{|\lambda|-2})\right)$ that is the normalized
character evaluated on the cycle class of transpositions. The profile $(2,1^{|\lambda|-2})$
is related to the so-called \textit{simple branch point}.

\br\label{Hurwitz-genus-formula} The quantity $d-\ell(\lambda)$ which is used in the definition (\ref{phi}) 
is called the \textit{length of permutation} with cycle structure $\lambda$ and will be denoted by $\ell^*(\lambda)$
(also called the \textit{colength} of the partition $\lambda$).
The colength enters the well-known Riemann-Hurwitz formula which relates the Euler characteristic 
$\textsc{e}$ of a base surface
to the Euler characteristic   $\textsc{e}'$  of it's $d$-branched cover:
\[
 \textsc{e}'- d\textsc{e}+\sum_{i} \ell^*(\Delta^{(i)})=0
\]
where the sum is over all branch points $z_i\,,i=1,2,\dots$ and where the ramification profiles are given by partitions 
$\Delta^i\,,i=1,2,\dots$.
\er

Let us introduce
\be\label{degree}
{\rm deg} \, \phi_\lambda(i)= i
\ee
According to (\ref{phi})
this degree is equal to the colength of ramification profiles, and due to Remark 
\ref{Hurwitz-genus-formula}
it defines the Euler characteristic $\e'$ of the covering surfaces.

From (\ref{tilde-phi}) we get
\be
{\rm deg}\,{\tilde \phi}_\lambda(i;\e) =i
\ee

In the notations of Theorem \ref{Theorem} we obtain
\bp
\[
Z_{\Sigma}(\rho;\Theta_1,\dots,\Theta_k)=\sum_{d\ge 0}\sum_{m\ge 0}\rho^m 
\sum_{\Delta^1,\dots,\Delta^k \in \Upsilon_d
\atop d=\ell(\Delta^1)=\cdots =\ell(\Delta^k)}
{\tilde{H}}^{\e_{i,m}}_\e(\Delta^1,\dots,\Delta^k;i,m) \prod_{v=1}^k\Theta^{\Delta^v}_v
\]
where 
\be\label{tilde-H}
 {\tilde{H}}^{\e_{i,m}}_\e(\Delta^1,\dots,\Delta^k;i,m):=\sum_{ \lambda \in  \Upsilon_d }
 \left[ {\tilde{\phi}}_\lambda(i;\e) \left(\phi_\lambda(1) \right)^m \right]
\varphi_\lambda(\Delta^1)\cdots 
\varphi_\lambda(\Delta^\f)\left( \frac{{\rm dim}\lambda}{|\lambda|!}\right)^{\e}
\ee
\ep
The coefficient in the square brackets is a polynomial function in  normalized characters $ \varphi_\lambda $ of
degree $ m + i $ (see (\ref{degree})).
Therefore, thanks to (\ref{Mednykh-Hurwitz}) this is a linear combination of
Hurwitz numbers, where the Euler characteristics of the base surface and its cover are equal, respectively, to
 $ \e $ and to
$ \e_{i, m} = d \e - \sum_{v = 1}^k \ell^*(\Delta^{(v)}) -m -i $. 
The difference with (\ref{Mednykh-Hurwitz})
comes from the coefficient in square brackets where $ (\phi(1))^m $ describes $ m $ extra simple
branch points, while $ {\tilde \phi} (i, \e) $ generates additional branch points
in number, not more than $ i $, with weights given by (\ref{tilde-phi}).

\section*{Acknowledgements}

The work of S.N. was partially supported by RFBR grant 20-01-00579
and MPIM.
The authors are grateful to S.Lando, M.Kazarian, J.Harnad, A.Morozov, A.Mironov, L.Chekhov and also to
Y.Marshall and V.Geogdjaev for various discussions.
We thank A.Gerasimov who paid our attention to \cite{Witten},
A.O. is grateful to A.Odzijewicz for his kind hospitality in Bialowezie and to  
E.Strahov, who turned his attention to independent Ginibre ensembles \cite{Ak1},\cite{S2},\cite{S1}.
A.O. was partially supported by V.E. Zakharov's scientific school
(Program for Support of Leading Scientific Schools, grant NS-3753.2014.2), by RFBR grant 18-01-00273a  and also 
by the Russian Academic Excellence Project `5-100'.

\end{document}